\documentclass[3p]{elsarticle}
\usepackage{amssymb}
\usepackage{amsfonts}
\usepackage{tikz}
\usetikzlibrary{decorations.pathmorphing,patterns,calc
,arrows,shapes,snakes,automata,backgrounds,petri
}

\newtheorem{theorem}{Theorem}[section]
\newtheorem{conjecture}[theorem]{Conjecture}

\newtheorem{lemma}[theorem]{Lemma}
\newtheorem{observation}[theorem]{Observation}

\newtheorem{claim}{Claim}

\newproof{proof}{Proof}

\begin{document}
\title{On the neighbour sum distinguishing index of planar graphs}

\author[LIRMM]{M. Bonamy\fnref{fn1}}
\ead{marthe.bonamy@lirmm.fr}

\author[agh]{J. Przyby{\l}o\fnref{fn2,fn3}}
\ead{jakubprz@agh.edu.pl}

\fntext[fn1]{Partly supported by the ANR Grant EGOS (2012-2015) 12 JS02 002 01}
\fntext[fn2]{Supported by the National Science Centre, Poland, grant no. 2014/13/B/ST1/01855.
}
\fntext[fn3]{Partly supported by the Polish Ministry of Science and Higher Education.}

\address[LIRMM]{LIRMM, Universit\'e Montpellier 2}
\address[agh]{AGH University of Science and Technology, al. A. Mickiewicza 30, 30-059 Krakow, Poland}

\begin{abstract} Let
$c$ be a proper edge colouring of a graph $G=(V,E)$ with
integers $1,2,\ldots,k$.
Then $k\geq \Delta(G)$, while by Vizing's theorem,
no more than $k=\Delta(G)+1$ is necessary for
constructing such $c$.
On the course of investigating irregularities in graphs,
it has been moreover
conjectured
that only slightly
larger $k$, i.e.,
$k=\Delta(G)+2$
enables enforcing additional strong feature of $c$,
namely
that it attributes distinct sums of incident colours
to adjacent vertices in $G$
if only
this graph
has no isolated edges
and is not isomorphic to $C_5$.
We prove the conjecture is valid
for planar graphs of sufficiently large
maximum degree.
In fact even stronger statement holds, as
the necessary number of colours stemming
from the result of Vizing
is proved to be sufficient for this family of graphs.
Specifically,
our main result states that every planar graph $G$ of maximum degree
at least $28$
which contains no isolated edges
admits a proper edge colouring
$c:E\to\{1,2,\ldots,\Delta(G)+1\}$
such that $\sum_{e\ni u}c(e)\neq \sum_{e\ni v}c(e)$ for every edge $uv$ of $G$.
\end{abstract}


\maketitle

\section{Introduction}
\emph{Every graph of order at least two
contains a pair of vertices of the same degree.}
This commonly known basic fact has brought forth a number of implicative questions.
In particular an issue of a possible definition of an \emph{irregular graph} was raised
by Chartrand, Erd\H{o}s and Oellermann in~\cite{ChartrandErdosOellermann}.
Note that with multiple edges admitted,
the solution seems obvious,
as
an \emph{irregular multigraph}
--
understood as a multigraph with pairwise distinct vertex degrees
--
exists for every order exceeding two.
As the same is far from being true in the case of
(simple) graphs,
Chartrand et al.~\cite{Chartrand} altered
towards measuring the `level of irregularity' of these instead.
Suppose that for a given graph $G=(V,E)$ we want to construct an irregular multigraph
of it via multiplying some of its edges.
The least $k$ so that we are able to do it
using at most $k$ copies of every edge
is known as the \emph{irregularity strength} of $G$ and denoted by $s(G)$.
Alternatively, one may reformulate
the same
in the language of (not necessarily proper) edge colourings
$c:E\to\{1,2,\ldots,k\}$,
assigning every edge an integer corresponding to its multiplicity in a desired multigraph,
where by
$$d_c(v)=\sum_{u\in N(v)}c(uv)$$
we shall denote so-called \emph{weighted degree} of $v\in V$.
The least $k$ so that such colouring exists attributing every vertex $v$ of $G$ a distinct
sum $d_c(v)$ of its incident colours is then equal to $s(G)$.
Note that $s(G)$ is well defined for all graphs
containing no isolated edges and at most one isolated vertex.
The irregularity strength was studied in numerous papers, e.g.~\cite{Aigner,Bohman_Kravitz,Lazebnik,
Faudree,Frieze,KalKarPf,Lehel,MajerskiPrzybylo2,Nierhoff,Przybylo,irreg_str2},
and was the cornerstone of many later graph invariants
and a new general direction in research on graphs, which might be referred to as
\emph{additive graph labelings},
or more generally -- \emph{vertex distinguishing graph colourings}.

One of the most intriguing questions
of the field concerns almost the same colouring problem as described
above, but with the global requirement $d_c(u)\neq d_c(v)$ for $u,v\in V$, $u\neq v$,
replaced with a local one, i.e., restricted only to the cases when $u$ and $v$ are adjacent
in $G$. We call $u$ and $v$ \emph{neighbours} then.
Karo\'nski \L uczak and Thomason~\cite{123KLT} conjectured that labels $1,2,3$
are sufficient to design a colouring meeting the local requirement
for every connected graph of order at least $3$,
and thus the problem
is commonly referred to
as
the \emph{1--2--3 Conjecture} in the literature.
Upon publication of the initial paper,
it was not even known if any finite set composed of initial positive integers
was sufficient for that goal.
The finite bound was settled and improved in~\cite{Louigi30,Louigi},
two intriguing articles including consequential
results reaching beyond this particular field.
See also~\cite{WangYu} and~\cite{KalKarPf_123}, where finally the set $\{1,2,3,4,5\}$
was proved to work, leaving us just `two steps' of the ultimate goal.

In this paper we focus on a correspondent of the \emph{1--2--3 Conjecture}
in the environment of \textbf{proper} edge colourings.
An assignment $c:E\to \{1,2,\ldots,k\}$ with $c(e)\neq c(e')$ for every pair of incident edges $e,e'\in E$
shall be called a \emph{proper edge $k$-colouring}.
Such colouring is said to be \emph{neighbour sum distinguishing}, or \emph{nsd} for short,
if for every edge $uv\in E$, there is no \emph{conflict} between $u$ and $v$,
i.e., $d_c(u)\neq d_c(v)$.
The least $k$ for which an nsd (edge) $k$-colouring of $G$ exists
is called the \emph{neighbour sum distinguishing index}, and denoted by $\chi'_{\sum}(G)$.
Note that this graph invariant is well defined
for all graphs without isolated edges.
We obviously have $\chi'_{\sum}(G)\geq \chi'(G)$ for these graphs then,
where by Vizing's theorem, $\chi'(G)$ equals the maximum degree of $G$, $\Delta(G)$, or $\Delta(G)+1$.
The following daring conjecture was on the other hand proposed
by Flandrin et al. in~\cite{FlandrinMPSW},
where it was also verified
for a few classical graph families, including, e.g.,
paths, cycles, complete
graphs, complete bipartite graphs and trees.
\begin{conjecture}\label{Flandrin_et_al_Conjecture}
If $G$ is a connected graph of order
at least three different from the cycle $C_5$, then
$\chi'_{\sum}(G) \leq \Delta(G) + 2$.
\end{conjecture}
In general it is known that this conjecture is asymptotically correct,
as confirmed by
the following probabilistic result of
Przyby{\l}o from~\cite{Przybylo_asym_optim}.
\begin{theorem}
\label{Th_sum_asymptotic}
If $G$ is a connected graph of maximum degree $\Delta\geq 2$, then $\chi'_{\sum}(G)\leq (1+o(1))\Delta$.
\end{theorem}
This was preceded by
other general upper bounds, involving the \emph{colouring number} of $G$, ${\rm col}(G)$,
defined as the least integer $k$ such that $G$ has a vertex enumeration in which each vertex
is preceded by fewer than $k$ of its neighbours (hence ${\rm col}(G)-1\leq\Delta(G)$).
In particular in \cite{Przybylo_CN_1} and \cite{Przybylo_CN_2}, the bounds
$\chi'_{\sum}(G)\leq 2\Delta(G)+{\rm col}(G)-1$ and
$\chi'_{\sum}(G)\leq \Delta(G)+3{\rm col}(G)-4$
(or even $\chi'_{\sum}(G)\leq \Delta(G)+3{\rm col}(G)-5$ for non-trees), resp.,
were proved for every graph $G$ containing no isolated edges
by means of algebraic tools
based on Combinatorial Nullstellensatz by Alon, see~\cite{Alon}.
The algebraic technique applied had also additional advantage,
as its characteristic provided
the same results in a more general list setting as well.
The second of these
also
yields upper bounds for $\chi'_{\sum}(G)$ of the form $\Delta(G)+const.$
for many classes of graphs with bounded colouring numbers
(cf. Conjecture~\ref{Flandrin_et_al_Conjecture}).
In particular it implies that
$\chi'_{\sum}(G)\leq \Delta(G)+13$ for every planar graph $G$ without an isolated edge.
Independently, planar graphs were also investigated in~\cite{DongWang_planar} and~\cite{WangChenWang_planar}, where
the bounds $\chi'_{\sum}(G)\leq\max\{2\Delta(G)$ $+1,25\}$ and $\chi'_{\sum}(G)\leq\max\{\Delta(G)+10,25\}$, resp., were proved for these graphs.
See also~\cite{DongWang_planar,WangYan_sum} for other results concerning $\chi'_{\sum}(G)$.
The main result of this paper, see Theorem~\ref{main_result_BP} below, not only implies
that Conjecture~\ref{Flandrin_et_al_Conjecture} is valid for
planar graphs
of maximum degree at least $28$,
but also strengthen it by assuring sufficiency of an upper
bound from Vizing's theorem to hold within our much more restrictive setting,
i.e., we prove that $\chi'_{\sum}(G)\leq \Delta(G)+1$ for these graphs.

It is worth mentioning that the concept investigated in this article was also inspired by
another central problem in the field of vertex distinguishing graph colourings.
The least integer $k$ so that a proper colouring $c:E\to\{1,2,\ldots,k\}$
exists such that the sets of colours incident with $u$ and $v$ are distinct for every edge $uv$ of $G$ is called
the \emph{neighbour set distinguishing index} (or \emph{adjacent strong chromatic index}) of $G$
and denoted by $\chi'_a(G)$.
Due to the properness of the colourings investigated, this is obviously a weaker requirement
than studied within the nsd colourings, and thus
$\chi'_a(G)\leq\chi'_{\sum}(G)$ for all graphs containing no isolated edges.
The neighbour set distinguishing index
was introduced by Zhang, Liu and Wang \cite{Zhang} in~2002 together with the following
challenging conjecture
--
constituting a weaker protoplast of Conjecture~\ref{Flandrin_et_al_Conjecture}
--
which gave another boost to the field, as it
triggered and inspired a large number of associated results.
\begin{conjecture}
\label{Zhang_Conjecture}
If $G$ is a connected graph of order
at least three different from the cycle $C_5$, then
$\chi'_a(G) \leq \Delta(G) + 2$.
\end{conjecture}
This conjecture was, e.g., verified by Balister et al.~\cite{BalGLS} for bipartite graphs and for graphs of maximum degree $3$, while Greenhill and Ruci\'nski proved it for almost all $4$-regular graphs (asymptotically almost surely), see~\cite{Rucinski_regular}. Recently it was also verified for planar graphs of maximum degree $\Delta\geq 12$ \cite{Hornak_planar} by means of the discharging method.
Independently, Bonamy, Bousquet and Hocquard~\cite{BonamyEtAl} improved the latter result
by proving an upper bound
$\chi'_a(G)\leq \Delta(G)+1$ for the same family of graphs.
The main result of this paper might thus be viewed at as a significant strengthening of this result
for planar graphs with maximum degree $\Delta\geq 28$.
Conjecture~\ref{Zhang_Conjecture} also holds for some families of graphs with bounded maximum average degree, ${\rm mad}(G)$,
see \cite{BonamyEtAl,HocqMont,WangWang} for details.
 In general it is known that $\chi'_a(G)\leq 3\Delta(G)$, see~\cite{Akbari},
and $\chi'_a(G)\leq \Delta(G)+O(\log \chi(G))$, see~\cite{BalGLS}.
The asymptotically best upper bound of the form $\chi'_a(G)\leq \Delta(G)+300$
was proved for every graph with no isolated edges and with maximum degree $\Delta>10^{20}$
by Hatami~\cite{Hatami}
by means of a multistage probabilistic construction.
One of the reasons the concept of sums is significantly harder to investigate than
similar problems focused on distinguishing merely by sets
is  that the probabilistic method
is much more unwieldy for application in `sum environment' due to the concentration of sums of independent random variables with a uniform distribution, and many other reasons.
In particular the probabilistic approach applied for proving Theorem~\ref{Th_sum_asymptotic}
is completely different
from the one used by Hatami, whose result could not be
modified or developed
towards
a stronger upper bound
for $\chi'_{\sum}(G)$.
Similarly, a list of new techniques
have been designed in order to tackle the case of planar graphs.
The
basic tool among this is the concept of dynamical list modifications introduced and used
several times throughout this paper.
Surprisingly,
this direction
has occurred to be easier in application, and more universal and effective
than
our initial approach based on
Alon's Combinatorial Nullstellensatz.

\section{Main result}
\begin{theorem}\label{main_result_BP}
Any planar graph $G$ with $\Delta(G) \geq 28$ and no isolated edge satisfies $\chi'_\Sigma(G) \leq \Delta(G)+1$.
\end{theorem}

For any graph $H$, set $n_i(H)=|\{v \in V(H) | d_H(v)=i\}$ for $i=1,2,\ldots$.
We say that the graph $H$ is \emph{smaller} than
a graph $H'$ if $|E(H)|<|E(H')|$ or $|E(H)|=|E(H')|$ and $(n_t(H),n_{t-1}(H),$ $\ldots,$ $n_2(H),$ $n_1(H))$ precedes
$(n_t(H'),n_{t-1}(H'),\ldots,n_2(H'),n_1(H'))$ with respect to the standard
lexicographic order, where $t=\max\{\Delta(H),\Delta(H')\}$.
We call a graph \emph{minimal} for a property when no smaller graph satisfies it. In the figures, black vertices have no other neighbours than those represented, while white vertices might have other neighbours as well as coincide with other white vertices. Numbers inside white vertices indicate the actual numbers of their neighbours.

We shall prove Theorem~\ref{main_result_BP} by contradiction.
Thus let from now on $k$ be an integer with $k \geq 28$ and $G=(V,E)$ be a minimal graph such that
$\Delta(G) \leq k$, $G$ is planar, has no isolated edges and $\chi'_\Sigma(G)>k+1$.

\section{Forbidden Configurations}

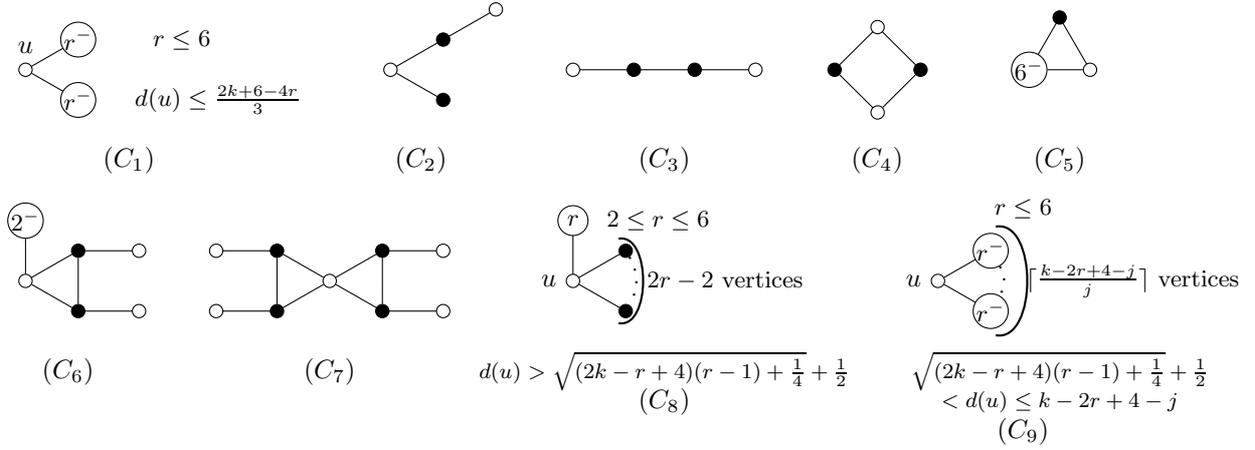
\begin{figure}[h]
\begin{center}
\centering
\begin{tikzpicture}[scale=0.8,auto]
\tikzstyle{w}=[draw,circle,fill=white,minimum size=5pt,inner sep=0pt]
\tikzstyle{b}=[draw,circle,fill=black,minimum size=5pt,inner sep=0pt]
\tikzstyle{t}=[rectangle,minimum size=5pt,inner sep=0pt]

\draw (0,0) node[w] (u) [label=90:$u$] {}
--++ (30:1cm) node[w] (v1) {\small $r^-$};

\draw (u)
--++ (-30:1cm) node[w] (v2) {\small $r^-$};

\draw (v1)
++(0:1.7cm) node[t] (t1) {\small $r \leq 6$};
\draw (v2)
++(0:2.3cm) node[t] (t1) {\small $d(u) \leq \frac{2k+6-4r}{3}$};
\draw (1.7,-1.5) node[t] (t1) {$(C_1)$};

\draw (6,0) node[w] (u) {}
--++ (30:1cm) node[b] (v1) {}
--++ (30:1cm) node[w] (v3) {};

\draw (u)
--++ (-30:1cm) node[b] (v2) {};

\draw (6.5,-1.5) node[t] (t1) {$(C_2)$};

\draw (9,0) node[w] (u) {}
--++ (0:1cm) node[b] (v1) {}
--++ (0:1cm) node[b] (v2) {}
--++ (0:1cm) node[w] (v3) {};

\draw (10.5,-1.5) node[t] (t1) {$(C_3)$};

\draw (13.3,0) node[b] (u) {}
--++ (45:1cm) node[w] (v1) {}
--++ (-45:1cm) node[b] (v2) {};

\draw (u)
--++ (-45:1cm) node[w] (v3) {};

\draw (v2) -- (v3);

\draw (14,-1.5) node[t] (t1) {$(C_4)$};

\draw (16.5,0) node[w] (u) {\small $6^-$}
--++ (60:1cm) node[b] (v1) {}
--++ (-60:1cm) node[w] (v2) {};

\draw (v2) -- (u);

\draw (17,-1.5) node[t] (t1) {$(C_5)$};

\draw (0,-3.5) node[w] (u) {}
--++ (30:1cm) node[b] (v1) {}
--++ (0:1cm) node[w] (v2) {};

\draw (u)
--++ (-30:1cm) node[b] (v3) {}
--++ (0:1cm) node[w] (v4) {};

\draw (u)
--++ (90:1cm) node[w] {\small $2^-$};

\draw (v1) -- (v3);

\draw (0.7,-5) node[t] (t1) {$(C_6)$};

\draw (5,-3.5) node[w] (u) {}
--++ (30:1cm) node[b] (v1) {}
--++ (0:1cm) node[w] (v2) {};

\draw (u)
--++ (-30:1cm) node[b] (v3) {}
--++ (0:1cm) node[w] (v4) {};

\draw (u)
--++ (30:-1cm) node[b] (v5) {}
--++ (0:-1cm) node[w] (v6) {};

\draw (u)
--++ (-30:-1cm) node[b] (v7) {}
--++ (0:-1cm) node[w] (v8) {};

\draw (v1) -- (v3);
\draw (v7) -- (v5);

\draw (5,-5) node[t] (t1) {$(C_7)$};

\draw(9,-3.5) node[w] (u) [label=left:$u$] {}
--++ (90:1cm) node[w] (v) {\small $\ r \ $};

\draw (u)
--++ (30:1cm) node[b] (v1) {};

\draw (u)
--++ (-30:1cm) node[b] (v3) {};

\draw (9.75,-3.5+0.7) edge [bend left=100,thick] node {} (9.75,-3.5-0.7);

\draw (v1) edge [bend left,loosely dotted,thick] node {} (v3);

\draw (v)
++(0:1.4cm) node[t] (t2) {\small $2 \leq r \leq 6$};

\draw (u)
++(0:2.5cm) node[t] (t1) {\small $2r-2$ vertices};

\draw (9+1.5,-5) node[t] (t3) {\footnotesize $d(u) > \sqrt{(2k-r+4)(r-1)+\frac{1}{4}}+\frac{1}{2}$};

\draw (10.5,-5.5) node[t] (t1) {$(C_8)$};

\draw(15,-3.5) node[w] (u) [label=left:$u$] {};

\draw (u)
--++ (30:1cm) node[w] (v1) {\small $r^-$};

\draw (u)
--++ (-30:1cm) node[w] (v3) {\small $r^-$};

\draw (9.75+6.2,-3.5+0.9) edge [bend left=100,thick] node {} (9.75+6.2,-3.5-0.9);

\draw (v1) edge [bend left,loosely dotted,thick] node {} (v3);

\draw (16.4,-2.3) node[t] (t2) {\small $r \leq 6$};

\draw (u)
++(0:3.2cm) node[t] (t1) {\small $\lceil\frac{k-2r+4-j}{j}\rceil$ vertices};

\draw (9+8,-5) node[t] (t3) {\footnotesize $\sqrt{(2k-r+4)(r-1)+\frac{1}{4}}+\frac{1}{2}$};
\draw (9+8,-5.5) node[t] (t3) {\footnotesize $<d(u) \leq k-2r+4-j$};

\draw (16.4,-6) node[t] (t1) {$(C_9)$};




\end{tikzpicture}
\caption{Forbidden configurations in $G$.}\label{FCG_fig}
\end{center}
\end{figure}

Below we define Configurations $(C_1)$ to $(C_9)$, see Figure~\ref{FCG_fig},
whose non-existence in $G$ shall be proven further.
A cycle of length $3$ in $G$ shall be called a \emph{triangle}.
(Note that a triangle does not need to form a face.)
A vertex of degree one in a graph
shall in turn be called a \emph{pendant vertex}.
\begin{enumerate}
\item $(C_1)$ is a vertex $u$ of degree at most $\frac{2k+6-4r}{3}$ adjacent to two vertices
      of degree at most $r\leq 6$ (where $\frac{2k+6-4r}{3} > 12$).
\item $(C_2)$ is a vertex $u$ adjacent to a vertex of degree $2$ and a vertex of degree $1$.
\item $(C_3)$ is two adjacent vertices of degree $2$.
\item $(C_4)$ is two vertices of degree $2$ adjacent to the same two vertices.
\item $(C_5)$ is a triangle containing both a vertex of degree $2$ and another vertex of degree at most $6$.
\item $(C_6)$ is a vertex $u$ adjacent to two adjacent vertices of degree $3$ and to a vertex of degree at most $2$.
\item $(C_7)$ is a vertex $u$ adjacent to two pairs of adjacent vertices of degree $3$.
\item $(C_8)$ is a vertex $u$ of degree $d(u)>\sqrt{(2k-r+4)(r-1)+\frac{1}{4}}+\frac{1}{2}$
adjacent to $2r-2$
pendant vertices and a vertex $v$ of degree $r\in [2,6]$.
\item $(C_9)$ is a vertex $u$ with $\sqrt{(2k-r+4)(r-1)+\frac{1}{4}}+\frac{1}{2}<d(u)\leq k-2r+4-j$  adjacent to $p\geq \frac{k-2r+4-j}{j}$ vertices of degree at most $r\leq 6$
    for some positive integer $j$.
\end{enumerate}

Before we show that ($C_1$)--($C_9$) do not appear in $G$,
we need to prove the following general technical lemma and observation first.
For sets $A\subset V$, $B\subset E$, by $G[A]$ we shall mean the \emph{graph induced} by the vertices of $A$ in $G$, while the graph obtained of $G$ by removing all edges in $B$ shall be denoted by $G-B$
(or $G-b$ if $B=\{b\}$). Moreover, for the sake of formulation of Lemma~\ref{claim:difsum_General} below,  by an nsd colouring of a graph containing isolated edges
we mean an nsd colouring of its components of order at least $3$ and arbitrary colouring
of its isolated edges.

\begin{lemma}\label{claim:difsum_General}
For a given
vertex
$u\in V$ and $p$ edges
$uv_1,uv_2,\ldots,uv_p$ incident with $u$
such that $\Delta(G[\{v_1,\ldots,v_p\}])\leq 1$,
$d(v_i)\leq 6$ and $d(u)\leq k-2d(v_i)+3$ for $i=1,\ldots,p$,
and for any nsd ($k+1)$-colouring $\alpha$ of $G - E(G[\{u,v_1,\ldots,v_p\}])$,
there are at least
$1+ p(k-d(u)+3)-2\sum_{i=1}^pd(v_i)$ different ways of extending $\alpha$
into a proper edge ($k+1)$-colouring of $G$ such that no two result in the same sum of colours on the $uv_i$'s,
and within each of this extensions there are no conflicts between adjacent vertices in $G$,
except possibly between $u$ and its neighbours outside $\{v_1,\ldots,v_p\}$
or
between $u$ and its neighbours in $\{v_1,\ldots,v_p\}$ but with no neighbour in this set.
\end{lemma}
\begin{proof}
Without loss of generality, we assume that
$v_{2t-1}v_{2t}\in E$ for $t=1,2,\ldots,l$,
where $l\leq\frac{p}{2}$ is some non-negative integer, and no other edge of $G$ except for these join
two vertices in $\{v_1,\ldots,v_p\}$.
Note that $d(v_1),
\ldots,d(v_{2l})\geq 2$ then.
Consider any nsd ($k+1)$-colouring $\alpha$ of $G - E(G[\{u,v_1,\ldots,v_p\}])$.
For every consecutive $j=1,\ldots,p$ (separately) we remove from
$\{1,\ldots,k+1\}$ the colours of
already coloured edges incident with $uv_j$ in $G$, and for $j>2l$ we additionally
remove each colour
using which on $uv_j$ would cause
conflict between $v_j$
and some of its neighbours other than $u$.
Denote the obtained list of the remaining available colours for $uv_j$ by $L_j$, $j=1,\ldots,p$.
Note that $l_j:=|L_j| \geq (k+1)-(d(u)-p)-(d(v_j)-2)=k-d(u)+p-d(v_j)+3$ for $j\leq 2l$
and $l_j:=|L_j| \geq (k+1)-(d(u)-p)-(d(v_j)-1)-(d(v_j)-1)=k-d(u)+p-2d(v_j)+3$ for $j> 2l$.
Let
$L_j=\{c_{j,i} : i=1,\ldots,l_j\}$
with $c_{j,1}<c_{j,2}<\ldots<c_{j,l_j}$ for $j=1,\ldots,p$.
We modify these lists dynamically as follows
(we denote by $l'_j$ the modified up-to-date number of colours in the list associated with $uv_j$):
\begin{itemize}
\item[(a)] for $i$ from $1$ to $p-1$ consecutively, we remove for every $j \geq i+1$ any $c_{j,m}=c_{i,1}$
      (where we mean, here and similarly further on, $c_{i,1}$ from the up-to-date list for $uv_i$)

     and if $i=2t-1$ for some $t\leq l$, we remove from the list associated with $uv_{2t}$
     a colour (if needed) $c^*$ such that setting $\alpha(uv_{2t})=c^*$ together with
     $\alpha(uv_{2t-1})=c_{2t-1,1}$
     would cause a conflict
     between $v_{2t-1}$ and $v_{2t}$ (regardless of the colour of $v_{2t-1}v_{2t}$);
\item[(b)] for $i$ from $p$ to $2$ consecutively, we remove for every $j \leq i-1$ any $c_{j,m}=c_{i,l'_i}$
           (where $c_{i,l'_i}\neq c_{j,1}$ for every $j \leq i-1$ by (a))

     and if $i=2t$ for some $t\leq l$, we remove from the list associated with $uv_{2t-1}$
     a colour $c^{**}$ such that setting $\alpha(uv_{2t-1})=c^{**}$ together with
     $\alpha(uv_{2t})=c_{2t,l'_{2t}}$
     would cause a conflict
     between $v_{2t-1}$ and $v_{2t}$
     (where $c^{**}\neq c_{2t-1,1}$ by (a)).
\end{itemize}
Note that by those operations, from the list of colours available for $uv_j$
we have removed at most $p$ colours
if $j\leq 2l$, or at most $p-1$ colours if $j>2l$, hence
$l'_j \geq
k-d(u)-2d(v_j)+4\geq 1$ for every $j$.
Moreover, it holds that\\$c_{1,1}+c_{2,1}+\ldots+c_{p-1,1}+c_{p,1}\\ <c_{1,1}+c_{2,1}+\ldots+c_{p-1,1}+c_{p,2} \\ < \ldots \\< c_{1,1}+c_{2,1}+\ldots+c_{p-1,1}+c_{p,l'_p} \\< c_{1,1}+c_{2,1}+\ldots+c_{p-1,2}+c_{p,l'_p} \\< \ldots \\< c_{1,l'_1}+c_{2,l'_2}+\ldots+c_{p-1,l'_{p-1}}+c_{p,1'_p}$.\\
These are at least
$\sum_{i=1}^p l'_i - (p-1) \geq pk-pd(u)-2\sum_{i=1}^pd(v_i)+4p-p+1 = 1+ p(k-d(u)+3)-2\sum_{i=1}^pd(v_i)$ different sums.
By the construction,
each of these consists of pairwise distinct colours
(thus
corresponds to a proper extension of the edge colouring on the $uv_i$'s)
with no possible conflicts between $v_i$ and its neighbours other than $u$ for $i> 2l$,
nor between $v_{2t-1}$ and $v_{2t}$, $t=1,\ldots,l$.
To complete the extensions of $\alpha$, for each of these we thus must only
choose colours for every edge $v_{2t-1}v_{2t}$
so that $v_{2t-1}$ and $v_{2t}$ are not in conflicts with their respective neighbours, $t=1,\ldots,l$.
We can always do that, since there are more than $4\times 5=20$ colours in $\{1,\ldots,k+1\}$.$~\blacksquare$
\end{proof}

One may note that Lemma~\ref{claim:difsum_General} above is optimal. 
It is sufficient to consider, e.g., a case when $l=0$ and
$d(v_j)=r$, $L_j=\{1,2,\ldots,(k+1)-(d(u)-p)-2(r-1)\}$ for every $j=1,\ldots,p$
to see that the quantity $1+ p(k-d(u)+3)-2\sum_{i=1}^pd(v_i)$
cannot be larger than claimed.

\begin{observation}\label{claim:conflict_Generalization}
In a proper edge $(k+1)$-colouring of $G$, if there is a conflict between two vertices $u$ and $v$
with $d(v)=r\leq 6$, then $d(u) \leq \sqrt{(2 k-r+4)(r-1)+\frac{1}{4}}+\frac{1}{2}$.
\end{observation}
\begin{proof}
The sum of colours of the edges incident to $v$ that are not incident to $u$ equals at most
$(k+1)+k+\ldots+(k-r+3) = \frac{1}{2}(2k-r+4)(r-1)$.
The sum of colours of the edges incident to $u$ but not to $v$ equals at least
$\sum_{i=1}^{d(u)-1} i = \frac{1}{2}d(u)\times(d(u)-1) = \frac{1}{2}[(d(u)-\frac{1}{2})^2-\frac{1}{4}]$. Thus, for a conflict to take place between $u$ and $v$,
we must have $(2k-r+4)(r-1) \geq (d(u)-\frac{1}{2})^2-\frac{1}{4}$, hence
$d(u) \leq \sqrt{(2 k-r+4)(r-1)+\frac{1}{4}}+\frac{1}{2}$.$~\blacksquare$
\end{proof}

\begin{lemma}
The graph $G$ does not contain any of Configurations $(C_1)$ to $(C_9)$.
\end{lemma}

\begin{proof}
By abuse of language, when we \emph{colour by minimality} a graph smaller than $G$, it actually means that we colour by minimality (fix some nsd $(k+1)$-colouring of) every connected component of it that is not an isolated edge, and we colour
every isolated edge with an arbitrary element from $\{1,\ldots,k+1\}$. For any vertex $v$, let $G \spadesuit \{v\}$ be the graph obtained from $G$ by replacing the vertex $v$ by $d(v)$ vertices of degree $1$, each adjacent to a different neighbour of $v$. Note that for any vertex $v$ with $d(v) \geq 2$, $G \spadesuit \{v\}$ is smaller than $G$.

\begin{claim}\label{claim:C_2_generalization}
The graph $G$ does not contain $(C_1)$.
\end{claim}

\begin{proof}
Assume by contradiction that there is a vertex $u$ with $d(u) \leq \frac{2k+6-4r}{3}$ in $G$ adjacent to two vertices $w$ and $x$ with $d(w), d(x) \leq r\leq 6$.
We consider two cases depending on whether $w$ and $x$ are adjacent.
\begin{itemize}
\item Assume $w$ and $x$ are adjacent.
    We colour by minimality $G - \{uw,ux,wx\}$.
    Since $d(u)\leq \frac{2k+6-4r}{3} \leq k-2r+3$ (as $r\leq 6$, $k\geq 28$),
    by Lemma~\ref{claim:difsum_General}, we can extend this colouring into a proper colouring
    in $1+ 2(k-d(u)+3)-2(d(w)+d(x)) \geq 2k-2d(u)+7-4r = d(u)+1+3\times(\frac{2k+6-4r}{3}-d(u)) \geq d(u)+1$ ways, each providing a distinct sum at $u$ and guaranteing no conflicts in $G$
    except possibly between $u$ and its neighbours other than $w$ and $x$.
    Since there are at most $d(u)-2$ such neighbours, at least one of these extensions yields
    an nsd $(k+1)$-colouring of $G$, a contradiction.
\item Assume $w$ and $x$ are not adjacent. We colour by minimality $G - \{uw,ux\}$.
    Similarly as above, by
    Lemma~\ref{claim:difsum_General},
    we can extend this colouring into a proper colouring in
    $1+ 2(k-d(u)+3)-2(d(w)+d(x)) \geq d(u)+1$ ways,
    each providing a distinct sum at $u$ and guaranteing no conflicts in $G$
    except possibly between $u$ and its neighbours, including $w$ and $x$ this time.
    However, in order to overcome this obstacle, it is sufficient to make
    one small alteration
    in the reasoning presented in the proof of Lemma~\ref{claim:difsum_General}.
    Namely, for $v_1=w$ and $v_2=x$,
    right before dynamical modifications of the lists $L_1,L_2$,
    we first additionally remove from $L_2$ the only colour (if it is contained in $L_2$)
    using which on $uv_2$ would cause conflict between $u$ and $v_1$ (regardless of the colour of $uv_1$),
    and we remove from $L_1$ the only colour
    using which on $uv_1$ would cause conflict between $u$ and $v_2$
    (consequently, $l'_j \geq k-d(u)-2d(v_j)+4-1\geq k - \frac{2k+6-4r}{3} -2r +3 \geq 1$ for $j=1,2$).
    This way we shorten each of the two lists by one additional element,
    and thus in total we obtain two less possible extensions (sums for $u$),
    i.e., at least $d(u)-1$ of these, but within each of these no conflict is possible
    between $u$ and $w$ nor between $u$ and $x$. Since there are $d(u)-2$ of the remaining
    neighbours of $u$, one of these extensions yields
    an nsd $(k+1)$-colouring of $G$, a contradiction.$~\blacksquare$
\end{itemize}
\end{proof}

\begin{claim}
The graph $G$ does not contain $(C_2)$.
\end{claim}
\begin{proof}
Assume there is a vertex $u$ adjacent to two vertices $v$ and $w$ with $d(v)=1$ and $d(w)=2$
in $G$. Let $x$ be the other neighbour of $w$. We colour by minimality $G \spadesuit \{w\}$.
By Configuration $(C_1)$, $d(u) > \frac{2k-2}{3} > \sqrt{2k+\frac{9}{4}}+\frac{1}{2}
$ as $k\geq 9$. Thus, by Observation~\ref{claim:conflict_Generalization},
there can be no conflict between $u$ and $w$. We consider two cases depending on $d(x)$.
\begin{itemize}
\item Assume $d(x) \leq \frac{k}{2}$. Since the colours of $uw$ and $uv$ can be switched,
      we may assume that the colour of $uw$ differs from the sum of colours on the edges
      incident with $x$
      other than $wx$.
      We then discolour the edge $wx$.
      There are at least $k+1-(\frac{k}{2}-1)-1$ colours available for $wx$
      (not used by its incident edges in $G$). To avoid conflicts between $x$ 
      and its neighbours,
      we eliminate additional at most $\frac{k}{2}-1$ 
      colours.
      Since $k+1-2\times(\frac{k}{2}-1)-1=2$, 
      there is a colour available for $wx$
      that induces an nsd $(k+1)$-colouring of $G$, a contradiction.
\item Assume $d(x) > \frac{k}{2} \geq \sqrt{2k+\frac{9}{4}}+\frac{1}{2}$ (as $k \geq 11$). Then by Observation~\ref{claim:conflict_Generalization}, there can be no conflict between $w$ and $x$. We switch if necessary the colours of $uv$ and $uw$ so that $uw$ and $wx$ are not coloured the same. This results in an nsd $(k+1)$-colouring of $G$, a contradiction.$~\blacksquare$
\end{itemize}
\end{proof}

\begin{claim}
The graph $G$ does not contain $(C_3)$.
\end{claim}
\begin{proof}
Assume $v,w$ are two adjacent vertices of degree $2$ in $G$, where $u\neq w$ is the other neighbour of $v$
and $x\neq v$ is the other neighbour of $w$.
If $u=x$, we colour by minimality $G - vw$. If $u \neq x$, we colour by minimality the graph $G$ where the edge $vw$ has been contracted. In both cases, the edges $uv$ and $wx$ are coloured differently.
In order to obtain an nsd $(k+1)$-colouring of $G$ it thus suffices to colour $vw$ so as to avoid conflicts between $u$ and $v$, and between $w$ and $x$. This is possible as there are more than $4$ colours available, a contradiction.$~\blacksquare$
\end{proof}

\begin{claim}
The graph $G$ does not contain $(C_4)$.
\end{claim}
\begin{proof}
Assume there are two vertices $v,x$ of degree $2$ adjacent to the same two vertices $u,w$ in $G$.
We colour by minimality $G \spadesuit \{v,x\}$ ($=(G \spadesuit \{v\}) \spadesuit \{x\}$). By Configuration $(C_1)$, $d(u),d(w)>
\frac{2k-2}{3}\geq \sqrt{2k+\frac{9}{4}}+\frac{1}{2}$. Thus, by
Observation~\ref{claim:conflict_Generalization}, there can be no conflict between $u$ or $w$ and $v$ or $x$. Then it suffices to construct a proper edge colouring of $uv,vw,wx$ and $ux$ in $G$ of the given one
in order to provide an nsd $(k+1)$-colouring of $G$.
This can be done by switching if necessary the colours of $uv$ and $ux$, a contradiction.$~\blacksquare$
\end{proof}

\begin{claim}
The graph $G$ does not contain $(C_5)$.
\end{claim}
\begin{proof}
Assume there is a triangle containing three vertices $u,v,w$ with $d(v)=2$ and $d(w) \leq 6$ in $G$.
We colour by minimality $G - vw$. If the colour of $uv$ is equal to the sum of colours incident with $w$, then the colour of $uv$ cannot be incident with $w$, thus we switch the colours of $uv$ and $uw$. Then there is no possible conflict between $v$ and $w$. It suffices to colour $vw$ (properly) so as to avoid conflicts of $v$ and $w$ with their neighbours other than $v$ and $w$.
There are more than $12$ colours, so this is possible, and thus we obtain
an nsd $(k+1)$-colouring of $G$, a contradiction.$~\blacksquare$
\end{proof}

\begin{claim}\label{claim:c7}
The graph $G$ does not contain $(C_6)$.
\end{claim}
\begin{proof}
Assume there is a vertex $u$ adjacent to two adjacent vertices $v,w$ of degree $3$
and to a vertex $x$ of degree at most $2$ in $G$. Let $y$ be the other neighbour of $x$ (if it exists).
We colour by minimality $G - vw$. By Configuration $(C_1)$,
$d(u)>
\frac{2k-6}{3}>\sqrt{2k+\frac{9}{4}}+\frac{1}{2}$ as $k\geq 12$. Thus, by
Observation~\ref{claim:conflict_Generalization},
there is no possible conflict between $u$ and $x$.

If $v$ and $w$ are incident to the same sum of colours, and the colour of $uv$ is incident to $w$ and symmetrically, then we switch the colour of $ux$ with that of $uv$ or $uw$. This is possible if $d(y)>\frac{k}{2}$ by choosing a colour different from that of $xy$ for $ux$ and by Observation~\ref{claim:conflict_Generalization}
(as $\frac{k}{2} \geq \sqrt{2k+\frac{9}{4}}+\frac{1}{2}$ for $k \geq 11$),
while if $d(y) \leq \frac{k}{2}$, we switch the colour of $ux$
so that the sum at $x$ is distinct from the sum at $y$.
If by accident this colour is the same as the colour of $xy$,
we change the colour of $xy$ so that it is distinct from
the colours of its at most $1+(\frac{k}{2}-1)$ incident edges
and so that $y$ is not in conflict with its at most $\frac{k}{2}-1$
neighbours other than $x$. This is feasible as we have $k+1$ colours available.
In the resulting colouring, $v$ and $w$ are not incident to the same sum of colours.

If the sum of colours incident to $v$ and the sum of colours incident to $w$ are equal, and the colour of $uv$ is not incident to $w$ (in our initial colouring), we switch the colours of $uv$ and $uw$. This results in a proper colouring where $v$ and $w$ are not incident to the same sum of colours.

When (finally) the sum of colours incident to $v$ is different from that of colours incident to $w$,
there obviously is an available
colour for $vw$ that raises no conflict between $v$ or $w$ and their respective neighbours,
and thus completes
an nsd $(k+1)$-colouring of $G$, a contradiction.$~\blacksquare$
\end{proof}

\begin{claim}
The graph $G$ does not contain $(C_7)$.
\end{claim}
\begin{proof}
Assume there is a vertex $u$ adjacent to two pairs $v_1w_1$ and $v_2w_2$ of adjacent vertices of degree $3$ in $G$. We colour by minimality $G - \{v_1w_1,v_2w_2\}$,
and we proceed similarly as in the proof of Claim~\ref{claim:c7}.
In particular, it is easy to see that we may always switch the colours of $uv_1, uw_1, uv_2$ and $uw_2$
(if necessary)
so that the colouring remains proper, the sum at $v_1$ is distinct from the sum at $w_1$
and the sum at $v_2$ is distinct from the sum at $w_2$.
Then we can easily complete the colouring by choosing available colours for $v_1w_1$ and $v_2w_2$
so that $v_1,w_1,v_2,w_2$ receive sums distinct from their corresponding neighbours, a contradiction.$~\blacksquare$
\end{proof}

\begin{claim}\label{claim:pendants_and_big}
The graph $G$ does not contain $(C_8)$.
\end{claim}
\begin{proof}
Assume by contradiction that
for some integer $r\in [2,6]$, there is a vertex
$u$ of degree $d(u)>\sqrt{(2k-r+4)(r-1)+\frac{1}{4}}+\frac{1}{2}$
adjacent to $2r-2$
pendant vertices and a vertex $v$ of degree $r$ in $G$.
We colour by minimality a graph obtained of $G$ by disjoining $uv$ from $v$,
or in other words by removing $uv$, then adding a new vertex $v'$ and joining it with $u$ by an edge.
There are at least $2r-1$ pendant vertices
adjacent with $u$ (corresponding to $2r-1$ edges coloured differently) in such graph.
Thus we may obtain an nsd colouring of $G$ by identifying one of these vertices with $v$,
since at most $r-1$ colours are already used by edges incident with $v$, and
at most $r-1$ further might be forbidden for fear of conflict between $v$ and its neighbours other than $u$
(while there can be no conflict between $v$ and $u$ by Observation~\ref{claim:conflict_Generalization}),
a contradiction.$~\blacksquare$
\end{proof}

\begin{claim}\label{claim:general_lemma}
The graph $G$ does not contain $(C_9)$.
\end{claim}

\begin{proof}
Assume $r,j$ are positive integers, $r\leq 6$, and suppose that $u$ is a vertex of degree $d$ in $G$, $\sqrt{(2k-r+4)(r-1)+\frac{1}{4}}+\frac{1}{2}<d\leq k-2r+4-j$,
adjacent with $p\geq \frac{k-2r+4-j}{j}$ vertices $v_1,\ldots,v_p$ of degree at most $r$ in $G$.
We colour by minimality the graph $G - E(G[\{u,v_1,\ldots,v_p\}])$.
Since $d(v_1),\ldots,d(v_p)\leq 6$, then
by
Configuration $(C_1)$,
$\Delta(G[\{v_1,\ldots,v_p\}])\leq 1$.
Thus, as $d\leq k-2r+4-j\leq k-2d(v_i)+3$ for $i=1,\ldots,p$,
by Lemma~\ref{claim:difsum_General}, there are at least
$1+ p(k-d+3)-2\sum_{i=1}^pd(v_i)\geq 1+ p(k-d+3)-2pr
= d-p+1 +p(k-2r+4) -d(p+1) \geq d-p+1 +p(k-2r+4) - (k-2r+4-j)(p+1)
= d-p+1 -(k-2r+4) + j(p+1) \geq d-p+1 -(k-2r+4) + j((\frac{k-2r+4}{j}-1)+1)
= d-p+1$
extensions of our colouring into a proper edge $(k+1)$-colouring of $G$ with no conflicts,
except possibly between $u$ and its neighbours,
each providing a different sum of the $uv_i$'s.
However,
by Observation~\ref{claim:conflict_Generalization} there can be no conflicts between $u$
and its neighbours $v_1,\ldots,v_p$. Since there are $d-p$ remaining neighbours of $u$,
at least one of our extensions yields an nsd $(k+1)$-colouring of $G$, a contradiction.$~\blacksquare$
\end{proof}

\flushright{$~\blacksquare$}
\end{proof}

\section{The graph $G$ is not planar}
In this section we shall prove that in fact $G$ cannot be planar
(and thus cannot be a counterexample to Theorem~\ref{main_result_BP}).

For this aim we analyse planar embeddings of some graphs.
For any planar mapping of a (planar) graph $H$,
the \emph{degree} of its face $f$, denoted $d_H(f)$ (or simply $d(f)$),
 is defined as the number of edges on
 the boundary walk (or walks) of $f$.
Note that the graphs investigated (e.g., $G$) do not have to be 2-connected.
Thus a given edge (or vertex) may appear more than once on a boundary walk (walks)
of one face,
contributing the corresponding number of times
to $d_H(f)$.
Moreover, even if an edge is incident with the same face ``from both sides'',
for convenience, in further comments we shall usually refer to it as to two different faces.
Analogously, if a vertex appears more than once on a boundary walk of some face,
each incidence of this vertex with this face shall be referred to as an incidence with a different
face.

\subsection{Trash}

\begin{figure}[t]
\begin{center}
\centering
\begin{tikzpicture}[scale=1,auto]
\tikzstyle{w}=[draw,circle,fill=white,minimum size=5pt,inner sep=0pt]
\tikzstyle{b}=[draw,circle,fill=black,minimum size=5pt,inner sep=0pt]
\tikzstyle{d}=[draw,densely dotted, thick, circle,minimum size=8pt,inner sep=0pt]
\tikzstyle{t}=[rectangle,minimum size=5pt,inner sep=0pt]

\draw (0,0) node[w] (u) {\small $7^+$}
--++ (90:1cm) node[b] (v) {};
\draw (v) node[d] (w) {};

\draw (u)
++(-90:0.7cm) node[t] (t) {$T_1$};

\draw (2,0) node[b] (u) {}
--++ (90:1cm) node[b] (v) {};

\draw (u)
--++ (30:1cm) node[w] (w1) {\small $7^+$};
\draw (u)
--++ (150:1cm) node[w] (w2) {\small $7^+$};
\draw (u) node[d] (w) {};
\draw (v) node[d] (w) {};

\draw (v) -- (w1);
\draw (v) -- (w2);

\draw (u)
++(-90:0.7cm) node[t] (t) {$T_2$};

\draw (5,1) node[b] (u) {};

\draw (u)
--++ (-30:1cm) node[w] (w1) {\small $7^+$};
\draw (u)
--++ (-150:1cm) node[w] (w2) {\small $7^+$};
\draw (u) node[d] (w) {};

\draw (w2) -- (w1);

\draw (u)
++(-90:1.7cm) node[t] (t) {$T_3$};

\draw (8,1) node[b] (u) {}
++(-90:1cm) node[w] (v) {};

\draw (u)
--++ (-30:1cm) node[w] (w1) {\small $7^+$};
\draw (u)
--++ (-150:1cm) node[w] (w2) {\small $7^+$};
\draw (u) node[d] (w) {};

\draw (v) -- (w1);
\draw (v) -- (w2);

\draw (v)
++(-90:0.7cm) node[t] (t) {$T_4$};




\end{tikzpicture}
\caption{Trash vertices (circled in dots).}\label{Trash_fig}
\end{center}
\end{figure}
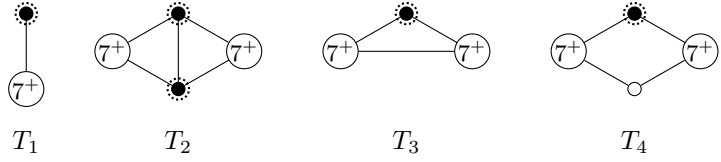

Consider any fixed planar mapping $\mathcal{M}$ of our graph $G$
(which makes up a minimal counterexample to Theorem~\ref{main_result_BP}).
In the following, for every $v\in V$,
by $d(v)$ we shall always mean $d_G(v)$.
We shall remove (choose) a special subset of vertices from $G$, which we shall call \emph{the trash} (see Figure~\ref{Trash_fig}).
To do that we proceed as follows performing the removals in the order listed below,
where on each stage we refer to the up-to-date reminiscence of our graph, denoted $G_t$,
and its embedding or faces
(but with vertex degrees referring to the original graph $G$). Set $G_t=G$.
\begin{enumerate}
\item First we remove every vertex $v$ with $d(v)=1$ from $G_t$. (Note that such $v$ must be incident with a vertex of degree greater than $\frac{k+2}{2}\geq 15\geq 7$ due to Configuration
    $(C_9)$ - it is sufficient to set $r=1$ and $j=\lceil\frac{k+2}{2}\rceil$ in it.)
    Denote the set of these vertices by $T_1$.
\item Second, we remove from (the obtained) $G_t$ every pair $u,v$ of adjacent vertices of degree $3$
($d(u)=3=d(v)$),
      sharing two common neighbours (which must both be of degree at least $17\geq 7$ due to ($C_1$)). Denote the set of all these removed vertices by $T_2$.
\item Third, remove every vertex $v$ with $d(v)=2$ contained in a face of degree $3$ in $G_t$.
      (Note that both neighbours of $v$ must be of degree at least $7$ due to ($C_5$).)
      Denote the set of these vertices by $T_3$.
\item Fourth, remove every vertex $v$ with $d(v)=2$ contained in a face of degree $4$ in $G_t$,
      and with both neighbours of degree at least $7$ (in $G$). Denote the set of such vertices by $T_4$.
\end{enumerate}
We denote the \emph{trash} set as $V_T=T_1\cup T_2\cup T_3\cup T_4$ and set $V'=V\smallsetminus V_T$. Then $G'=G[V']$
is the graph obtained of $G$ after all the four operations above.
Its resulting planar embedding shall be denoted $\mathcal{M}'$.

\begin{observation}\label{obs:noT1234inG'}
Note that $G'$ does not contain a vertex $v$ with $d(v)=1$, nor a vertex $v$ of degree
$3$ (in $G$) with three neighbours $u,x,y$ such that $d(u)=3$ and $u$ is adjacent to $x$ and $y$,
nor (by ($C_4$)) a vertex $v$ with $d(v)=2$ contained in a face of degree $3$ of $\mathcal{M}'$,
nor a vertex $v$ with $d(v)=2$ contained in a face of degree $4$ of $\mathcal{M}'$, with both neighbours of degree at least $7$ (in $G$).
\end{observation}
To see that it is sufficient to note that due to Configuration ($C_4$), by removing the vertices of $T_3$,
we cannot create in $G_t$ any new vertices `of type $T_3$' (i.e., vertices of degree $2$ in $G$,
contained in in a face of degree $3$ of $G_t$), and then similarly, again due to ($C_4$),
by removing the vertices of $T_4$,
we cannot create in $G_t$ any new vertices `of type $T_3$ or $T_4$'.

Moreover, since all neighbours of the vertices in $V_T$ are of degree at least $7$
(with exception for the pairs from $T_2$), we obtain the following observation, 
where
for a given set $A\subset V$ and a vertex $v\in V$, by $d_A(v)$ we shall mean the number of neighbours
of $v$ (from $G$) in $A$.

\begin{observation}\label{obs:degree6degree6}
For every vertex $v\in V'$ with $d(v)\leq 6$, $d_{V'}(v)=d(v)$.
\end{observation}

\subsection{Euler's formula}

\begin{observation}\label{obs:pl}
Let $(V',V_T)$ be a partition (defined above) of the vertices of our planar graph $G$ with planar embedding
$\mathcal{M}$, and $\mathcal{M}'$ being the induced embedding of its subgraph $G'=G[V']$.
Denote the set of faces of $\mathcal{M}'$ by $F'$.
Assign to each vertex $v$ of $G$ a weight of $d_G(v)-6$ and to each face $f$ of $\mathcal{M}'$ a weight of $2d(f)-6$. It is not possible to discharge the weight over the graph in such a way that all the vertices and faces of $\mathcal{M}'$ have non-negative weights, while every vertex $v$ of $V_T$ has a weight of at least $d_G(v)+d_{V'}(v)-6$.\end{observation}

\begin{proof}
Suppose it is possible. Then,
\begin{eqnarray}
  \sum_{v \in V'\cup V_T}(d_G(v)-6)+\sum_{f \in F'}(2d(f)-6) &\geq& \sum_{v \in V_T}(d_G(v)+d_{V'}(v)-6)\nonumber\\
  \sum_{v \in V'}(d_{V'}(v)+d_{V_T}(v)-6)+\sum_{v \in V_T}(d_G(v)-6)+\sum_{f \in F'}(2d(f)-6) &\geq& \sum_{v \in V_T}(d_G(v)+d_{V'}(v)-6)\nonumber\\
  \sum_{v \in V'}(d_{V'}(v)-6)+\sum_{v \in V'}d_{V_T}(v)+\sum_{v \in V_T}(d_{G}(v)-6)+\sum_{f \in F'}(2d(f)-6) &\geq& \sum_{v \in V_T}(d_G(v)+d_{V'}(v)-6)\nonumber\\
  \sum_{v \in V'}(d_{V'}(v)-6)+\sum_{v \in V_T}d_{V'}(v)+\sum_{v \in V_T}(d_{G}(v)-6)+\sum_{f \in F'}(2d(f)-6) &\geq& \sum_{v \in V_T}(d_G(v)+d_{V'}(v)-6)\nonumber\\
  \sum_{v \in V'}(d_{V'}(v)-6)+\sum_{v \in V_T}(d_{G}(v)+d_{V'}(v)-6)+\sum_{f \in F'}(2d(f)-6) &\geq& \sum_{v \in V_T}(d_G(v)+d_{V'}(v)-6)\nonumber\\
  \sum_{v \in V'}(d_{V'}(v)-6)+\sum_{f \in F'}(2d(f)-6) &\geq& 0,\nonumber
\end{eqnarray}
which is in contradiction with Euler's formula applied to $\mathcal{M}'$.$~\blacksquare$
\end{proof}

Note that if such a contradiction can be reached in $\mathcal{M}'$, then it could also be reached in $\mathcal{M}$. Thus Observation~\ref{obs:pl} does not yield a more powerful result than based straightforwardly on Euler's formula, 
but it helps to decrease the number of rules in the discharging process and simplifies its analysis.
%

\subsection{Discharging rules}

We assign to each vertex $v$ in $G$ a weight of
$d(v)-6$, and to each face $f$ of $\mathcal{M'}$ a weight of
$2d(f)-6$.

Below we define rules of discharging,
which,
as shall be exhibited, result in redistribution of the weights inconsistent with
Observation~\ref{obs:pl}.

Within all the following rules, the degree of a vertex shall refer to its degree in $G$,
while the faces and their degrees correspond to the embedding $\mathcal{M'}$ of $G'$.

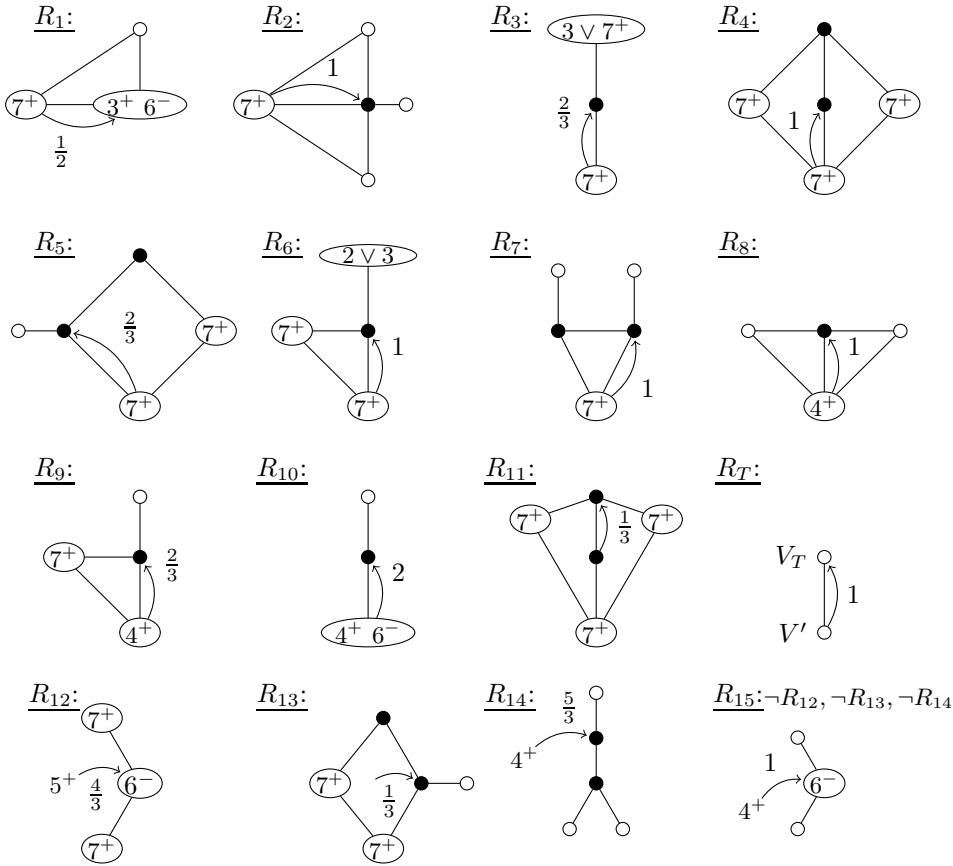
\begin{figure}[h]
\begin{center}
\begin{tikzpicture}[scale=1,auto]
\tikzstyle{w}=[draw,ellipse,fill=white,minimum size=5pt,inner sep=0pt]
\tikzstyle{b}=[draw,circle,fill=black,minimum size=5pt,inner sep=0pt]
\tikzstyle{d}=[draw,densely dotted, thick, circle,minimum size=8pt,inner sep=0pt]
\tikzstyle{t}=[rectangle,minimum size=5pt,inner sep=0pt]

\draw (0,0) node[w] (u) {\small $3^+ \ 6^-$} 
--++(0:-1.5cm) node[w] (v) {\small $7^+$};

\draw (u)
--++ (90:1cm) node[w] (w) {};
\draw (v) -- (w);
\draw (u) edge [bend left,pre] node {$\frac{1}{2}$} (v);
\draw (u)
++(-45:-1.6cm) node[t] (t1) {\underline{$R_1$:}};

\draw (3,0) node[b] (u) {}
--++(0:-1.5cm) node[w] (v) {\small $7^+$};

\draw (u)
--++ (90:1cm) node[w] (w) {};
\draw (u)
--++ (-90:1cm) node[w] (w2) {};
\draw (u)
--++ (0:0.5cm) node[w] (w3) {};
\draw (v) -- (w);
\draw (v) -- (w2);
\draw (v) edge [bend left,post] node {$1$} (u);
\draw (u)
++(-45:-1.6cm) node[t] (t1) {\underline{$R_2$:}};

\draw (6,-1) node[w] (v) {\small $7^+$}
--++(90:1cm) node[b] (u) {}
--++(90:1cm) node[w] (w) {\small $3 \vee 7^+$};

\draw (v) edge [bend left,post] node {$\frac{2}{3}$} (u);
\draw (u)
++(-45:-1.6cm) node[t] (t1) {\underline{$R_3$:}};

\draw (9,-1) node[w] (v) {\small $7^+$}
--++(90:1cm) node[b] (u) {}
--++(90:1cm) node[b] (w) {}
--++(-45:1.4cm) node[w] (x1) {\small $7^+$};

\draw (w)
--++(45:-1.4cm) node[w] (x2) {\small $7^+$};
\draw (v) -- (x1);
\draw (v) -- (x2);

\draw (v) edge [bend left,post] node {$1$} (u);
\draw (u)
++(-45:-1.6cm) node[t] (t1) {\underline{$R_4$:}};


\draw (0,-3) node[t] (u) {}
++(0:1cm) node[w] (v1) {\small $7^+$};

\draw (u)
++(-90:1cm) node[w] (v4) {\small $7^+$};

\draw (u)
++(90:1cm) node[b] (v2) {};

\draw (u)
++(0:-1cm) node[b] (v3) {}
--++(0:-0.6cm) node[w] (w) {};

\foreach \i in {1,...,4}
{\pgfmathtruncatemacro{\j}{mod(\i,4)+1}
    \draw (v\i) -- (v\j);}
\draw (v3) edge [bend left,pre] node {$\frac{2}{3}$} (v4);
\draw (u)
++(-45:-1.6cm) node[t] (t1) {\underline{$R_5$:}};

\draw (3,-3) node[b] (u) {}
--++(90:1cm) node[w] (v1) {\small $\ 2 \vee 3 \ $};

\draw (u)
--++(-90:1cm) node[w] (v2) {\small $7^+$};

\draw (u)
--++(0:-1cm) node[w] (v3) {\small $7^+$};

    \draw (v2) -- (v3);
\draw (u) edge [bend left,pre] node {$1$} (v2);
\draw (u)
++(-45:-1.6cm) node[t] (t1) {\underline{$R_6$:}};

\draw (6,-3) node[t] (u) {}
++(0:0.5cm) node[b] (v1) {};

\draw (u)
++(0:-0.5cm) node[b] (v2) {};

\draw (u)
++(90:-1cm) node[w] (v3) {\small $7^+$};

\draw (v1)
--++(90:0.8cm) node[w] (w1) {};
\draw (v2)
--++(90:0.8cm) node[w] (w2) {};

   \draw (v1) -- (v2);
    \draw (v2) -- (v3);
    \draw (v1) -- (v3);
\draw (v1) edge [bend left,pre] node {$1$} (v3);
\draw (u)
++(-45:-1.6cm) node[t] (t1) {\underline{$R_7$:}};

\draw (9,-3) node[b] (u) {}
--++(0:1cm) node[w] (v1) {};

\draw (u)
--++(0:-1cm) node[w] (v2) {};

\draw (u)
--++(90:-1cm) node[w] (v3) {\small $4^+$};

    \draw (v2) -- (v3);
    \draw (v1) -- (v3);
\draw (u) edge [bend left,pre] node {$1$} (v3);
\draw (u)
++(-45:-1.6cm) node[t] (t1) {\underline{$R_8$:}};


\draw (0,-6) node[b] (u) {}
--++(90:0.8cm) node[w] (v1) {};

\draw (u)
--++(0:-1cm) node[w] (v2) {\small $7^+$};

\draw (u)
--++(90:-1cm) node[w] (v3) {\small $4^+$};

\draw (v2) -- (v3);
\draw (u) edge [bend left,pre] node {$\frac{2}{3}$} (v3);
\draw (u)
++(-45:-1.6cm) node[t] (t1) {\underline{$R_9$:}};

\draw (3,-7) node[w] (v) {\small $4^+ \ 6^-$}
--++(90:1cm) node[b] (u) {}
--++(90:0.8cm) node[w] (w) {};

\draw (u) edge [bend left,pre] node {$2$} (v);
\draw (u)
++(-45:-1.6cm) node[t] (t1) {\underline{$R_{10}$:}};

\draw (6,-7) node[w] (v) {\small $7^+$}
--++(90:1cm) node[b] (u) {}
--++(90:0.8cm) node[b] (w) {};

\draw(u)
++(30:1cm) node[w] (x1) {\small $7^+$};
\draw(u)
++(150:1cm) node[w] (x2) {\small $7^+$};
\draw (x1) -- (w);
\draw (x1) -- (v);
\draw (x2) -- (w);
\draw (x2) -- (v);

\draw (w) edge [bend left,pre] node {$\frac{1}{3}$} (u);
\draw (u)
++(-45:-1.6cm) node[t] (t1) {\underline{$R_{11}$:}};

\draw (9,-6) node[w] (u) [label=left:$V_T$] {}
--++(90:-1cm) node[w] (v) [label=left:$V'$] {};

\draw (u) edge [bend left,pre] node {$1$} (v);
\draw (u)
++(-45:-1.6cm) node[t] (t1) {\underline{$R_{T}$:}};


\draw (0,-9) node[w] (u) {$6^-$}
--++(-60:-1cm) node[w] (v1) {\small $7^+$};

\draw (u)
--++(60:-1cm) node[w] (v2) {\small $7^+$};

\draw (u)
++(0:-1cm) node[t] (f) {\small $5^+$};

\draw (u) edge [bend right,pre] node[pos=0.2] {$\frac{4}{3}$} (f);
\draw (u)
++(-45:-1.6cm) node[t] (t1) {\underline{$R_{12}$:}};

\draw (3.7,-9) node[b] (u) {}
--++(0:0.6cm) node[w] (v1) {};

\draw (u)
--++(60:-1cm) node[w] (v2) {\small $7^+$};
\draw (u)
--++(-60:-1cm) node[b] (v3) {};
\draw (u)
++(0:-1.2cm) node[w] (w) {\small $7^+$};
\draw (v2) -- (w);
\draw (v3) -- (w);

\draw (u)
++(0:-0.7cm) node[t] (f) {};

\draw (u) edge [bend right,pre] node[pos=0.2] {$\frac{1}{3}$} (f);
\draw (f)
++(-45:-1.6cm) node[t] (t1) {\underline{$R_{13}$:}};

\draw (6,-9) node[b] (u) {}
--++(90:0.6cm) node[b] (v1) {}
--++(90:0.6cm) node[w] (v2) {};

\draw (u)
--++(60:-0.7cm) node[w] (v4) {};
\draw (u)
--++(-60:0.7cm) node[w] (v3) {};

\draw (u)
++(-20:-1cm) node[t] (f) {\small $4^+$};

\draw (f) edge [bend left,post] node[pos=0.65] {$\frac{5}{3}$} (v1);
\draw (u)
++(-45:-1.6cm) node[t] (t1) {\underline{$R_{14}$:}};

\draw (9,-9) node[w] (u) {\small $6^-$}
--++(60:-0.7cm) node[w] (v2) {};

\draw (u)
--++(-60:-0.7cm) node[w] (v4) {};

\draw (u)
++(20:-1cm) node[t] (f) {\small $4^+$};

\draw (f) edge [bend left,post] node[pos=0.65] {$1$} (u);
\draw (u)
++(-45:-1.6cm) node[t] (t1) {\underline{$R_{15}$:}};
\draw (u)
++(68:1.2cm) node[t] (t2) {\small $\neg R_{12}, \neg R_{13}, \neg R_{14}$};




\end{tikzpicture}
\caption{Discharging Rules.}\label{DR_fig}
\end{center}
\end{figure}

First we design $11$ discharging rules between vertices of $V'$ \emph{exclusively} (see Figure~\ref{DR_fig}):\\

For any vertex $u\in V'$ of degree at least $7$,
\begin{itemize}
\item $R_1$ is when $u$ is adjacent to a vertex $v$ with $3\leq d(v) \leq 6$ and the edge $uv$ belongs to at least one face of degree $3$. Then $u$ gives $\frac{1}{2}$ to $v$ (or more if one of the following rules applies to $uv$).
\item $R_2$ is when $u$ is adjacent to a vertex $v$ with $d(v)=4$ and the edge $uv$ belongs to two faces of degree $3$. Then $u$ gives a total of $1$ to $v$ (that is, an additional $\frac{1}{2}$
    adding up to $\frac{1}{2}$ from $R_1$).
\item $R_3$ is when $u$ is adjacent to a vertex $v$ with $d(v)=2$ such that the other neighbour of $v$ is of degree $3$ or of degree at least $7$. Then $u$ gives $\frac{2}{3}$ to $v$ (or $1$ if $R_4$ applies to $uv$).
\item $R_4$ is when $u$ is incident with two faces $(u,v,w,x)$ and $(u,v,w,x')$
     such that $d(v)=2$, $d(w)=3$ and $d(x),d(x')\geq 7$. Then $u$ gives a total of
     $1$ to $v$.
\item $R_5$ is when $u$ is incident with a face $(u,v,w,x)$ with $d(v)=3$, $d(w)=2$ and $d(x)\geq 7$.
      Then $v$ gives a total of $\frac{2}{3}$ to $v$ (or $1$ if $R_6$ applies to $uv$).
\item $R_6$ is when $u$ is adjacent to a vertex $v$ with $d(v)=3$ which is adjacent to
      vertices $w,x$ such that $d(w)\geq 7$, $d(x)\in\{2,3\}$ and $(u,v,w)$ is a face.
      Then $u$ gives a total of $1$ to $v$.
\item $R_7$ is when $u$ is adjacent to a vertex $v$ with $d(v)=3$ which
    is adjacent to a vertex $w$ with $d(w)=3$ such that $(u,v,w)$ is a face.
    Then $u$ gives a total of $1$ to $v$.\\
\end{itemize}

For any vertex $u\in V'$ of degree at least $4$,
\begin{itemize}
\item $R_8$ is when $u$ is adjacent to a vertex $v$ with $d(v)=3$
    and the edge $uv$ belongs to two faces of degree $3$. Then $u$ gives a total of $1$ to $v$.
\item $R_9$ is when $u$ is adjacent to a vertex $v$ with $d(v)=3$ and there is a vertex $w$ with $d(w) \geq 7$
    such that $(u,v,w)$ is a face. Then $u$ gives a total of $\frac{2}{3}$ to $v$
    (or $1$ if $R_6$, $R_7$ or $R_8$ applies to $uv$).\\
\end{itemize}

For any vertex $u\in V'$ of degree at least $4$ and at most $6$,
\begin{itemize}
\item $R_{10}$ is when $u$ is adjacent to a vertex $v$ with $d(v)=2$. Then $u$ gives $2$ to $v$.\\
\end{itemize}

For any vertex $u\in V'$ of degree $2$,
\begin{itemize}
\item $R_{11}$ is when $u$ is incident with two faces $(w,u,v,x)$ and $(w,u,v,x')$
     such that $d(v)=3$ and $d(w),d(x),d(x')\geq 7$. Then $u$ gives $\frac{1}{3}$ to $v$.\\
\end{itemize}

For any face $f\in \mathcal{M'}$ of degree at least $5$,
\begin{itemize}
\item $R_{12}$ is when $f$ is incident with a vertex $v$ with $2\leq d(v)\leq 6$
    whose two neighbours $w_1, w_2$ such that $w_1,v,w_2$ are consecutive vertices on the boundary walk of $f$
    are of degree at least $7$.
    Then $f$ gives $\frac{4}{3}$ to $v$.\\
\end{itemize}

For any face $f\in \mathcal{M'}$ of degree $4$,
\begin{itemize}
\item $R_{13}$ is when $f$ is of the form $(u,v,w,x)$ with $d(u)=2$, $d(v)=3$ and $d(w),d(x)\geq 7$.
      Then $f$ gives $\frac{1}{3}$ to $v$.\\
\end{itemize}

For any face $f\in \mathcal{M'}$ of degree at least $4$,
\begin{itemize}
\item $R_{14}$ is when $f$ is incident with two adjacent vertices $v, w$ with $d(v)=2$ and $d(w)=3$. Then $f$ gives $\frac{5}{3}$ to $v$.
\item $R_{15}$ is when $f$ is incident with a vertex $v$ with $2\leq d(v)\leq 6$
    and none of $R_{12}$, $R_{13}$ and $R_{14}$ applies to $f$ and $v$. Then $f$ gives $1$ to $v$.\\
\end{itemize}

Finally, we design a rule $R_T$ to deal with the trash:

For any vertex $u$ in $V'$,
\begin{itemize}
\item $R_T$ is when $u$ is adjacent to a vertex $v$ in $V_T$. Then $u$ gives $1$ to $v$.
\end{itemize}

\subsection{Discharging balance}

\begin{lemma}\label{lemma:discharging_ballance}
 After the discharging,
all the vertices and faces of $\mathcal{M'}$ have non-negative final weights,
while every vertex $v$ of $V_T$ has a weight of at least
$d(v)+d_{V'}(v)-6$.
\end{lemma}

\begin{proof}
Let $f$ be a face of $\mathcal{M'}$.
Note that
by $R_{12}, R_{13}, R_{14}$ and $R_{15}$, only vertices of \emph{small} degrees, i.e., degrees at most 6 might receive some weight from $f$.
\begin{itemize}
\item Assume $d(f)=3$. Face $f$ has an initial weight of $0$, and no rule applies, so it has the final weight of $0$.
\item Assume $d(f)=4$. Face $f$ has an initial weight of $2$.
       By Configuration $(C_1)$, $f$ gives a weight to at most $2$ vertices.
       If
       every such vertex receives at most $1$ from $f$, then
       $f$ has a non-negative final weight.
       It thus remains to consider the case when $R_{14}$ applies.
       But then, by Rule $R_{13}$, $f$ gives away all together $\frac{5}{3}+\frac{1}{3}$,
       and thus has the final weight of $0$.
\item Assume $d(f)\geq 5$. Face $f$ has an initial weight of $2d(f)-6$.
      By Configuration ($C_1$), $f$ gives weight to at most $d(f)-2$ vertices.
      But $f$ gives away at most $\frac{4}{3}$ \emph{on average} to each such vertex,
      since even if some vertex, say $v$, receives $\frac{5}{3}$ from $f$ via Rule $R_{14}$,
      then the neighbour of degree $3$ of $v$ receives $1$ from $f$ via Rule $R_{15}$
      (while there cannot be more than two consecutive vertices receiving a weight from $f$
      on its boundary walk by ($C_1$)).
      Thus $f$ has the final weight of at least $(2d(f)-6)-\frac{4}{3}(d(f)-2)=\frac{2}{3}d(f)-\frac{10}{3}\geq 0$.
\end{itemize}

Let $v$ be a vertex of $G$.
Since by Rule $R_T$ the final weight of every vertex $v\in V_T$ equals
$d(v)+d_{V'}(v)-6$, it is sufficient to consider
$v\in V'$, hence $d(v)\geq 2$.
Recall that by Observation~\ref{obs:degree6degree6}, if $d(v)\leq 6$, then $d(v)=d_{V'}(v)$, i.e., $v$ has no neighbours in $V_T$
(and hence does not give any weight away via Rule $R_T$).
We consider different cases depending on the degree of $v$.
\begin{itemize}
\item Assume $d(v)=2$. According to our rules,
      $v$ gives no weight away to any other vertex
      in all cases,
      except when Rule $R_{11}$ applies.
      Therefore, since the initial weight of $v$ equals $-4$,
      we need to show that it receives altogether
      at least $4$ from its neighbours and incident faces in all these cases.
      By Observation~\ref{obs:noT1234inG'}, $v$ cannot belong to a face of degree $3$
      (of $\mathcal{M}'$, or simply $G'$).
      Thus by $R_{12}$, $R_{14}$ and $R_{15}$, $v$ receives at least $2\times 1$
      from its incident faces. Denote the neighbours of $v$ by $u_1$, $u_2$, where $d(u_1)\leq d(u_2)$,
      and note that if $4\leq d(u_1)\leq 6$ (hence Rule $R_{11}$ does not apply to $v$),
      then $v$ has a non-negative final weight by Rule $R_{10}$.
      Assume that it is otherwise then.
      By Configuration ($C_3$),
      $v$ has no neighbours of degree $2$
      (and no neighbours of degree $1$ by Observation~\ref{obs:degree6degree6})
      either.
      Consider the following cases.
      \begin{itemize}
      \item Assume $d(u_1)\geq 7$. Then $d(u_2)\geq 7$, and hence $v$ receives $2\times \frac{2}{3}$
            from $u_1$ and $u_2$ by Rule $R_3$. Moreover, since $v\in V'$, hence $v\notin T_4$, then
            both faces (of $\mathcal{M}'$) incident with $v$ must be of degree at least $5$ 
            (see Observation~\ref{obs:noT1234inG'}),
            and thus give $2\times \frac{4}{3}$ to $v$ via Rule $R_{12}$.
            Since $R_{11}$ does not apply, $v$ has a non-negative final weight in this case.
      \item Assume $d(u_1)=3$ and $v$ gives away $\frac{1}{3}$ to $u_1$ via Rule $R_{11}$.
            Then $v$ receives $1$ from $u_2$ by Rule $R_4$ and $2\times \frac{5}{3}$
            by $R_{14}$ from its incident faces. Thus $v$ has a non-negative final weight.
      \item Assume $d(u_1)=3$ and $v$ does not give away $\frac{1}{3}$ to $u_1$ via Rule $R_{11}$.
           Then $d(u_2)\geq 7$ by ($C_1$), and thus $v$ receives $\frac{2}{3}$ from $u_2$ by Rule $R_3$. As $v$ additionally receives $2\times \frac{5}{3}$ via Rule $R_{14}$, it has a non-negative final weight in this case as well.
      \end{itemize}
\item Assume $d(v)=3$. According to our rules, $v$ gives no weight away.
      Therefore, since its initial weight equals $-3$, we need to show that $v$ receives altogether
      at least $3$ from its neighbours and incident faces. Denote the neighbours of $v$ by $u_1$, $u_2$ and $u_3$.
      Consider the following cases.
      \begin{itemize}
      \item Assume $v$ receives $\frac{1}{3}$ via Rule $R_{13}$ from (at least) one of its incident faces. Without loss of generality, assume that $u_1$ and $u_2$ belong to this face, with $d(u_1)=2$
          and $d(u_2)\geq 7$. Note that by ($C_1$), $d(u_3)\geq 7$.
          \begin{itemize}
          \item Suppose additionally that $u_1$ gives $\frac{1}{3}$ to $v$ via Rule $R_{11}$.
                Then $v$ receives also $\frac{1}{3}$ from the face $(u_1,v,u_3,\ldots)$ by Rule $R_{13}$
                (hence still lacks $2$ to a non-negative weight).
                Moreover, if $u_2,v,u_3$ form a face of degree $3$ in $G'$, then $v$ also receives
                $2\times 1$ from $u_2$ and $u_3$ via Rule $R_6$.
                Otherwise, $v$ receives at least $1$ from the face $(u_2,v,u_3,\ldots)$ by Rule
                $R_{12}$ or $R_{15}$ and $2\times \frac{2}{3}$ from $u_2$ and $u_3$ by Rule $R_5$.
                In both cases $v$ has a non-negative final weight.
          \item Suppose that $u_1$ does not give $\frac{1}{3}$ to $v$ via Rule $R_{11}$. Then
                the face $(u_1,v,u_3,\ldots)$ has degree at least $5$
                (e.g. by ($C_5$))
                and hence gives $1$ to $v$ by Rule $R_{15}$ (thus $v$ still lacks $\frac{5}{3}$ to a non-negative weight).
                Moreover, if $u_2,v,u_3$ form a face of degree $3$ in $G'$, then $v$ also receives
                $2\times 1$ from $u_2$ and $u_3$ via Rule $R_6$.
                Otherwise, $v$ receives at least $1$ from the face $(u_2,v,u_3,\ldots)$ by Rule
                $R_{12}$ or $R_{15}$ and $\frac{2}{3}$ from $u_2$ by Rule $R_5$.
                In both cases $v$ has a non-negative final weight.
          \end{itemize}
     \item Assume $v$ does not receive $\frac{1}{3}$ via Rule $R_{13}$ from any of its incident faces.
           Consider the following subcases.
           \begin{itemize}
           \item $v$ is not incident to any face of degree $3$ in $G'$.
                 Then $v$ receives at least $3\times 1$ from its incident faces via
                 Rule $R_{12}$ or $R_{15}$, and thus has a non-negative final weight.
           \item $v$ is incident with exactly $1$ face of degree $3$ in $G'$.
                 W.l.o.g., we may assume that it is the face
                 $(u_1,v,u_2)$ with $d(u_1)\leq d(u_2)$. Since $u_1\notin T_3$, then $d(u_1)\geq 3$,
                 while by Configuration ($C_1$), $d(u_2)\geq 7$.
                 Hence $v$ receives $1$ from $u_2$ via Rule $R_7$ if $d(u_1)=3$,
                 or otherwise, at least $\frac{1}{2}$ from $u_2$ by Rule $R_1$, $R_6$ or $R_9$
                 and at least $\frac{2}{3}$ from $u_1$ by Rule $R_6$ or $R_9$.
                 Thus $v$ has a non-negative final weight, as in both cases it additionally receives
                 no less than $2\times 1$ from its two incident faces of degree at least $4$ via
                 Rule $R_{12}$ or $R_{15}$.
           \item $v$ is incident with exactly $2$ faces of degree $3$ in $G'$.
                 Since $u_i\notin T_3$, then $d(u_i)\geq 3$ for $i=1,2,3$.
                 W.l.o.g., we may assume that the face
                 $(u_2,v,u_3,\ldots)$ has degree at least $4$, and thus gives at least
                 $1$ to $v$ via Rule $R_{12}$ or $R_{15}$.
                 Since $u_1\notin T_2$, then $d(u_1)\geq 4$, and thus $u_1$ gives $1$
                 to $v$ by Rule $R_8$.
                 Moreover, if $u_3$ (or symmetrically $u_2$) is of degree $3$,
                 and thus $u_1$ and $u_2$ ($u_3$) have degrees at least $7$ by Configuration ($C_1$),
                 then $u_2$ ($u_3$, resp.) gives $1$ to $v$ by Rule $R_6$.
                 Otherwise, $d(u_2),d(u_3)\geq 4$ and by ($C_1$) at most one of the vertices $u_1,u_2,u_3$
                 may have its degree smaller than $7$ (i.e., in $\{4,5,6\}$).
                 But then $u_2$ and $u_3$ give at least $2\times \frac{1}{2}$ to $v$
                 by $R_1$ or $R_9$.
                 In both cases $v$ has a non-negative final weight.
           \item $v$ is incident with $3$ faces of degree $3$ in $G'$.
                 Since $u_i\notin T_2$, then $d(u_i)\geq 4$ for $i=1,2,3$.
                 Therefore, $u_1,u_2,u_3$ give $3\times 1$ to $v$
                 via Rule $R_8$, and thus $v$ has the final weight of $0$.
           \end{itemize}
\end{itemize}
\item Assume $d(v)=4$.
      Denote the consecutive neighbours of $v$ by $u_1,u_2,u_3,u_4$,
      and assume $u_1$ has minimum degree among these. By Configuration ($C_1$), $d(u_i)\geq 7$
      for $i=2,3,4$.
      If $vu_3$ is incident with two faces of degree $3$,
      then by $R_1$ and $R_2$, $v$ receives at least $\frac{1}{2}+1+ \frac{1}{2}$ from $u_2, u_3$ and $u_4$. If $vu_3$ is incident with exactly one face of degree $3$, say $(u_2,v,u_3)$,
      then $v$ receives at least $\frac{1}{2}+\frac{1}{2}$ from $u_2$ and $u_3$ by $R_1$ or $R_2$, and at least $1$ from the face $(u_3,v,u_4,\ldots)$ by $R_{12}$ or $R_{15}$.
      If $vu_3$ is not incident with a face of degree $3$,
      then $v$ receives at least $1$ from each of the two faces $(u_2,v,u_3,\ldots)$ and $(u_3,v,u_4,\ldots)$ by $R_{12}$ or $R_{15}$.
      So, in all three cases, $v$ receives at least $2$. Since $v$ has an initial weight of $-2$, it only remains to ensure that vertex $v$ does not give away (to $u_1$) more than it additionally receives. We consider three cases depending of $d(u_1)$.
\begin{itemize}
\item Assume $d(u_1)=2$. Then $v$ gives $2$ to $u_1$ by $R_{10}$. Since $u_1\notin T_3$,
      the edge $vu_1$ belongs to two faces $f_1$ and $f_2$ of degree at least $4$. Thus $v$ receives (additionally) $1+1$ from $f_1$ and $f_2$ by $R_{15}$, and hence has a non-negative final weight. \item Assume $d(u_1)=3$. Then $v$ might give at most $1$ to $u_1$ by $R_8$ or $R_9$.
      In such a case, either the edge $vu_1$ belongs to a face of degree at least $4$ and $v$ receives an additional $1$ from it by $R_{15}$, or $vu_1$ belongs to
      two faces of degree $3$ and thus $v$ receives at least additional (not taken into account previously)
      $2 \times \frac{1}{2}$
      from $u_2$ and $u_4$ by $R_1$ or $R_2$. In both cases
      $v$ has a non-negative final weight.
\item Assume $d(u_1) \geq 4$. Then $v$ gives nothing away to $u_1$ and thus has a non-negative final weight.
\end{itemize}
\item Assume $5 \leq d(v) \leq 6$.
      Denote the consecutive neighbours of $v$ by $u_1,u_2,\ldots,u_l$, $5\leq l\leq 6$,
      and assume $u_1$ has minimum degree among these. By Configuration ($C_1$), $d(u_i)\geq 7$
      for $i=2,3,\ldots,l$.
      Vertex $v$ has an initial weight of at least $-1$. If the face $(u_3,v,u_4,\ldots)$
      is of degree greater than $3$,
      then it gives at least $1$ to $v$
      by Rule $R_{12}$ or $R_{15}$. Otherwise, $v$ receives $2\times \frac{1}{2}$ from $u_3$ and $u_4$
      by Rule $R_1$. In both cases it is thus sufficient to show that $v$ receives at least as much
      from its remaining neighbours and incident faces as it might give away to $u_1$.
      \begin{itemize}
      \item Assume $d(u_1)=2$. Then $v$ gives $2$ to $u_1$ by $R_{10}$. Since $u_1\notin T_3$,
        the edge $vu_1$ belongs to two faces $f_1$ and $f_2$ of degree at least $4$. Thus $v$ receives (additionally) $1+1$ from $f_1$ and $f_2$ by $R_{15}$, and hence has a non-negative final weight.
      \item Assume $d(u_1)=3$. Then $v$ might give at most $1$ to $u_1$ by $R_8$ or $R_9$.
        In such a case, either at least one of the faces $(u_2,v,u_3,\ldots)$ or $(u_4,v,u_5,\ldots)$
        is of degree greater than $3$, and thus gives (additionally) $1$ to $v$ by Rule $R_{12}$ or $R_{15}$,
        or otherwise, $v$ receives additionally $2\times \frac{1}{2}$ from $u_2$ and $u_5$ via Rule $R_1$.
        In both cases
        $v$ has a non-negative final weight.
      \item Assume $d(u_1) \geq 4$. Then $v$ gives nothing away to $u_1$ and thus has a non-negative final weight.
      \end{itemize}
\item Assume $7\leq d(v) \leq \frac{2k-18}{3}$.
      By
      Configuration $(C_1)$,
      $v$ is adjacent (in $G$) with at most one vertex of degree at most $6$, and thus
      gives at most $1$ away (as Rule $R_{10}$ does not apply to vertices with degree $\geq 7$).
      Since its initial weight was equal to $d(v)-6\geq 1$,
      it retains a non-negative final weight.
\item Assume $\frac{2k-18}{3}< d(v) \leq k-6$.
      First note that if $v$ has at least $12$ neighbours of degree at least $5$,
      then $v$ gives away at most $\frac{1}{2}$ to each of them by Rule $R_1$
      and at most $1$ to every of the rest of its neighbours via the remaining rules,
      hence in total at most
      $\frac{1}{2}\cdot 12 +(d(v)-12)\cdot 1 = d(v)-6$.
      As the initial weight of $v$ is equal to $d(v)-6$,
      it retains a non-negative final weight then.
      It remains to show that $v$ has at most $d(v)-12$ neighbours
      of degree at most $4$ in this case.
      \begin{itemize}
      \item Suppose that $\frac{2k-18}{3}<d(v)\leq\frac{2k-10}{3}$. Then by
       Configuration $(C_1)$,
            $v$ has at most $1$ neighbour of degree at most $4$.
            At the same time, as $k\geq 28$, we have $1\leq \lceil\frac{2k-18}{3}\rceil-12\leq d(v) -12$.

      \item Suppose that
      $\frac{2k-10}{3}<d(v)\leq k-6$.
           Then $d(v)> \sqrt{(2k-4+4)(4-1)+\frac{1}{4}}+\frac{1}{2}$ (as $k\geq 24$),
           and thus we may use Configuration $(C_9)$
           to analyse neighbours of $v$ of degree at most $4$.
      We consider the following subcases:
           \begin{itemize}
           \item Assume $\frac{2k-10}{3}< d(v)\leq k-9$. Since $d(v)\leq k-2\cdot 4+4-5$,
                 by
                 Configuration $(C_9)$, $v$ has at most
                 $\lceil\frac{k-9}{5}\rceil-1
                 \leq \lceil\frac{2k-10}{3}\rceil-12 \leq d(v)-12$ neighbours of degree at most $4$
                 (as $k\geq 27$).
           \item Assume $d(v)\in\{k-8,k-7\}$. Since $d(v)\leq k-2\cdot 4+4-3$,
                 by
                 Configuration $(C_9)$, $v$ has at most
                 $\lceil\frac{k-7}{3}\rceil-1
                 \leq k-8-12\leq d(v)-12$ neighbours of degree at most $4$ (as $k\geq 25$).
           \item Assume $d(v)=k-6$. Since $d(v)\leq k-2\cdot 4+4-2$,
                 by
                 Configuration $(C_9)$, $v$ has at most
                 $\lceil\frac{k-6}{2}\rceil-1
                 \leq k-6-12 = d(v)-12$ neighbours of degree at most $4$ (as $k\geq 28$).
           \end{itemize}
     \end{itemize}
\item Assume $d(v)=k-5$. Then $d(v)> \sqrt{(2k-3+4)(3-1)+\frac{1}{4}}+\frac{1}{2}$ (as $k\geq 13$).
      Since at the same time, $d(v)\leq k-2\cdot 3+4-3$, by
      Configuration $(C_9)$,
      $v$ has
      at most $\lceil\frac{k-5}{3}\rceil-1$
      neighbours of degree at most $3$. Hence $v$ has (in $G$) at least
      $k-5-(\lceil\frac{k-5}{3}\rceil-1)\geq 16$
      neighbours of degree at least $4$ (which are vertices of $G'$ as well).
      Let $A$ be the set of these neighbours $u$ of $v$ of degree exactly $4$
      for which
      $uv$ is contained in two
      faces of degree $3$ in $G'$, hence each such $u$ receives $1$ from $v$ via Rule $R_2$.
      Set $a=|A|$.
      Let $S$ be the set of
      neighbours $u$ of $v$ of degree at least $7$, hence each such $u$ receives $0$ from $v$.
      Set $s=|S|$.
      Note that since the initial weight of $v$ was equal to $d(v)-6$,
      then its final weight is non-negative if $s\geq 6$.
      Hence assume that $s\leq 5$.
      The remaining neighbours of $v$ of degree at least $4$ (which do not belong to $A$ nor to $S$)
      we denote by $R$.
      Note that each of these might receive at most $\frac{1}{2}$ from $v$ by Rule $R_1$,
      and $|R|\geq 16-a-s$.
      Since $v$ gives away at most $1$ to each neighbour of degree at most $3$,
      then its final weight equals at least
      $(d(v)-6)-(d(v)-|A|-|S|-|R|)-|A|-\frac{1}{2}|R| = -6+s+\frac{1}{2}|R| \geq
      -6+s+\frac{1}{2}(16-a-s) = 2+\frac{1}{2}(s-a)$.
      By
      Configuration $(C_1)$, for each vertex from $A$, one of the two faces of degree $3$ in $G'$ incident with it and with $v$ must be incident with a vertex from $S$.
      Therefore, since in turn a vertex from $S$ cannot be
      incident with more than two such faces,
      then $a\leq 2s$.
      Hence, the final weight of $v$ equals at least
      $2+\frac{1}{2}(s-2s)$, and thus (since $s\leq 5$) might be negative only if $s=5$ and $a=10$.
      But then, since $d_{V'}(v)> 15$, (by ($C_1$))
      there must exist at least two vertices $x_1,x_2\in V'\smallsetminus (A\cup S)$
      each of which belongs to a face of degree $3$ in $G'$
      (hence $d(x_1),d(x_2)\geq 3$ by Observation~\ref{obs:noT1234inG'})
      incident with $v$ and some vertex in $A$.
      Thus each of $x_1,x_2$ might receive
      at most $\frac{1}{2}$ from $v$ via Rule $R_1$. Indeed, otherwise, by ($C_1$), Rule $R_8$ or $R_9$ would have to apply to $vx_1$ or $vx_2$.
      Hence, again by ($C_1$), $vx_1$ or $vx_2$ would have to be incident with
      a face of degree $3$ in $G'$ including a vertex from $S$. This is however impossible as
      $a=10=2s$ implies that for every neighbour $u$ of $v$ with degree at least $7$,
      $uv$ belongs to exactly two faces of $G'$ - each of the form $(u,v,w)$ with $w\in A$.
      This (coupled with $|S|=5$) guarantees
      that $v$ has a non-negative final weight also in this case.
\item Assume $d(v)\geq k-4$. Then $d(v)\geq 24$.
      Let $u_0,u_1,\ldots,u_{d'-1}$ be the consecutive
      neighbours of $v$ in $G'$ (i.e., such that for all $i$, the vertices $u_i,v,u_{i-1}$, with indices understood modulo $d'$, are always consecutive vertices on a boundary walk of some face in $G'$). Denote the set
      of these by $N'_v$.
      Analogously, let $w_0,w_1,\ldots,w_{d-1}$ be the consecutive
      neighbours of $v$ in $G$, and denote the set of these by $N_v$. (Thus $|N_v|=d=d(v)$ and $|N'_v|=d'=d_{V'}(v)$.)
      For every $i$, we shall write that $u_i$ ($w_i$) \emph{precedes} $u_{i+1}$ in $G'$ ($w_{i+1}$ in $G$)
      and \emph{succeeds} $u_{i-1}$ in $G'$ ($w_{i-1}$ in $G$).
      \begin{itemize}
      \item Assume $v$ is not adjacent with a pendant vertex nor with two adjacent vertices of degree $3$ (in $G$).
       We shall show that $v$ gives away a total weight of at most $\frac{3}{4}d(v)\leq d(v)-6$
       to its neighbours $w_i$ in $G$.
       In order to
       do that we shall partition its neighbours
       into small subsets
       (each consisting of consecutive elements from $N_v$), called \emph{blocks},
       each of which shall be consistent with our goal, i.e., shall receive a total
       weight of at most $\frac{3}{4}$ of its cardinality from $v$.
       For the sake of our construction we shall also distinguish so-called \emph{half-blocks}
       (which shall be later paired into full blocks).

       Note that by ($C_5$) every neighbour $x$ of $v$ which belongs to $T_3$ must be succeeded or preceded
       by a neighbour $u$ of $v$ of degree at least $7$ in $G$,
       where $(u,v,x)$ is a face of degree $3$ in $G-T_1-T_2$
       (even in $G-T_1$ in our case, as $v$ is assumed not to be adjacent with a pendant vertex nor with two adjacent vertices of degree $3$).
       On the other hand, for every neighbour $u$ of $v$ of degree at least $7$,
       by ($C_4$), the edge $uv$ may belong to at most one such face.
       In order to simplify further notation, we make two modifications concerning the neighbourhood of $v$ in $G$. First, for every neighbour $u$ of $v$ of degree at least $7$ which does not belong
       to a face of the form $(u,v,x)$ with $d(x)=2$ in $G-T_1$,
       we introduce a new vertex $x$ and join it by edges with $u$ and $v$ so that
       $uv$ belongs to exactly one face of this type.
       Then every such pair of (consecutive in $G$) neighbours $u$ and $x$ (added or already having existed) of $v$ we call a \emph{pair of twins}.
       (Note that as every
       such $x$ `belongs to $T_3$' according to our definitions then,
       and thus shall be regarded as receiving $1$ from $v$
       via Rule $R_T$,
       then if we are able to
          prove that $v$ gives away a total weight of at most $\frac{3}{4}$ of its (new)
          degree, then it does the same with respect to its degree in the original graph $G$).
      Second, for the sake of defining the half-blocks, we locally and temporarily rearrange
          the weight distribution given away by $v$, not changing the total weight it gives away,
          i.e., we shall assume that $v$ gives away
          $\frac{1}{2}$ and $\frac{1}{2}$, instead of $1$ and $0$,
          to both vertices in every pair of twins
          (where each of the two twins shall be placed in one of the half-blocks making up together
          a full block at the end).

          At the beginning of our construction no block is defined,
          and we start by
          setting each twin as
          a 1-element half-block. These half-blocks might however be extended further on.
          Note that now by Configuration ($C_1$), every neighbour of $v$ in $N'_v$
          receiving at least $\frac{2}{3}$
          from $v$, except
          for those to which $R_3$ applies,
          must be preceded
          or succeeded by a twin in $G$ (cf. $R_2$, $R_4$, $R_5$, $R_6$, $R_8$ and $R_9$).
          We incorporate each such neighbour of $v$ into a half-block
          formed of one of such (at most two) its corresponding twins.
          As every twin is succeeded or preceded by a twin in $G$, then each half-block consists of
          at most two vertices afterwards, while its elements receive at most $(\frac{1}{2}+1)/2=\frac{3}{4}$ on average
          from $v$.

          Then there is just one kind of neighbours receiving more than $\frac{3}{4}$, namely $1$
          from $v$ in $G$ (which do not belong to any block or half-block yet),
          i.e., those in $T_4$.
          Consider such vertex $w$ ($d(w)=2$) and let $x\in N'_v$ be a non-neighbour of $w$
          such that $(w,v,x,y)$ is a face in $G-T_1-T_2-T_3$ for some vertex $y$ of degree at least $7$.
          Note that $d(x)\geq 3$ by Configuration ($C_4$).
          We proceed as follows with the consecutive vertices $w$ of this type
          subject to the features of their corresponding $x$'s
          (note that
          $x$ cannot receive $1$ from $v$, cf. Rules $R_2$,
          $R_6$, $R_7$ and $R_8$),
          while for more than one $x$ corresponding to $w$, we arbitrarily choose any of these.
          \begin{itemize}
          \item
          If $x$ belongs
          to a pair of twins,
          then we join $w$ with this twin of this pair which directly
          precedes or succeeds $w$ in $G$
          into a new half-block (note that such twin could not have yet been used in any extended, two-element block). Again, its elements receive no more than $\frac{3}{4}$ on average
          from $v$.
          \item If $x$ receives $\frac{2}{3}$ from $v$ (via Rule $R_5$ or $R_9$),
          then we include $w$ into the half-block to which $x$
          already belongs (making of it a half-block whose elements receive $(1+\frac{2}{3}+\frac{1}{2})/3<\frac{3}{4}$ on average from $v$).
          \item If $x$ receives $\frac{1}{2}$  from $v$ (via Rule $R_1$),
          then we make a block of $w$ and $x$ (since by Rule $R_1$, $x$ must belong
          to a face of degree $3$ including $v$ in $G'$, the vertex $x$ will not belong to two such,
          nor any other blocks simultaneously at this point).
          \item Finally, if $x$ receives $0$ from $v$ (where $x$ is not a twin,
          and hence $d(x)\leq 6$),
          then we make of $w$ and $x$ a block. Moreover, if $x$ belongs to two blocks of such type (this is the only such situation
          we admitted thus far, since until this subcase,
          no neighbour receiving $0$ from $v$ in our locally modified graph
          has been
          included in any block or half-block),
          then we unify these two blocks into one three-element block.
          Either way
          the elements of every block constructed in this step shall receive
          at most $\frac{2}{3}$ on average from $v$.
          \end{itemize}
          To complete our construction we join the corresponding pairs of half-blocks into full blocks
          and then we make 1-element blocks of all neighbours of $v$ in $G$ which do not yet belong to any block (each of these receive at most $\frac{2}{3}$ from $v$ by our construction).
          Such partition of $N_v$ fulfills our requirements,
          and thus $v$ has a non-negative final weight.
      \item Assume $v$ is adjacent with two adjacent vertices of degree $3$. Then by Configuration $(C_7)$ it cannot be adjacent with another pair of such vertices nor with any vertex of degree at most $2$ due to Configuration $(C_6)$.
          Thus $v$ might give at most
          $1$ to each of these two vertices,
          and no other neighbour of $v$ may belong in $V_T$ (hence they all belong in $G'$).
          Note that analogously as above, every other (except for these two) neighbour of $v$ receiving more than $\frac{2}{3}$,
          namely $1$ from it must be preceded or succeeded in $G'$ by a neighbour of degree at least $7$, receiving $0$ from $v$. Therefore $v$ gives away at most $2+\frac{2}{3}(d(v)-2)\leq d(v)-6$, and thus
          retains a non-negative final weight.
      \item Assume $v$ is adjacent with $p$ pendant vertices, $p\geq 1$. Then by Configurations $(C_2)$ and $(C_6)$, $v$ is not adjacent to vertices of degree $2$ (in particular those in $T_3$ and $T_4$)
          nor with a pair of adjacent vertices of degree $3$ (including those in $T_2$).
          \begin{itemize}
          \item Suppose $p\leq 6$. Then $d_{V'}(v)\geq 18$ and analogously as in the case above,
                $v$ gives away at most $\frac{2}{3}$ on average to its neighbours in $G'$. Hence $v$
                retains a non-negative final weight.
          \item Assume $p>6$. Then by
                Configuration $(C_8)$, $v$ cannot be adjacent with
                vertices of degrees $2,3,4$. Note also that
                $p<\frac{d(v)}{2}$ by Configuration $(C_9)$
                (it is sufficient to set $r=1$ and $j=2$ for $d(v)=k$ or $j=3$ for $k-4\leq d(v)\leq k-1$ in it),
                and hence $v$ is incident with
                more than $\frac{d(v)}{2}\geq 12$ neighbours of degree at least $5$, each of which receives at most $\frac{1}{2}$ from $v$ by Rule $R_1$. Thus $v$ has a non-negative final weight.$~\blacksquare$
          \end{itemize}
      \end{itemize}
\end{itemize}
\end{proof}

\subsection{Proof of Theorem~\ref{main_result_BP}}
By Lemma~\ref{lemma:discharging_ballance}, we obtain a contradiction with Observation~\ref{obs:pl}.
Thus $G$ cannot be planar, and therefore is not a counterexample to Theorem~\ref{main_result_BP}.
Hence no counterexample exists and the thesis follows.$~\blacksquare$

\end{document}